\documentclass[printer]{rQUF2e}

\usepackage{epstopdf}% To incorporate .eps illustrations using PDFLaTeX, etc.

\theoremstyle{plain}

\theoremstyle{definition}

\theoremstyle{remark}

\usepackage{scalerel}
\usepackage{tikz}
\usetikzlibrary{svg.path}

\definecolor{orcidlogocol}{HTML}{A6CE39}
\tikzset{
  orcidlogo/.pic={
    \fill[orcidlogocol] svg{M256,128c0,70.7-57.3,128-128,128C57.3,256,0,198.7,0,128C0,57.3,57.3,0,128,0C198.7,0,256,57.3,256,128z};
    \fill[white] svg{M86.3,186.2H70.9V79.1h15.4v48.4V186.2z}
                 svg{M108.9,79.1h41.6c39.6,0,57,28.3,57,53.6c0,27.5-21.5,53.6-56.8,53.6h-41.8V79.1z M124.3,172.4h24.5c34.9,0,42.9-26.5,42.9-39.7c0-21.5-13.7-39.7-43.7-39.7h-23.7V172.4z}
                 svg{M88.7,56.8c0,5.5-4.5,10.1-10.1,10.1c-5.6,0-10.1-4.6-10.1-10.1c0-5.6,4.5-10.1,10.1-10.1C84.2,46.7,88.7,51.3,88.7,56.8z};
  }
}

\newcommand\orcidicon[1]{\href{https://orcid.org/#1}{\mbox{\scalerel*{
\begin{tikzpicture}[yscale=-1,transform shape]
\pic{orcidlogo};
\end{tikzpicture}
}{|}}}}

\usepackage{hyperref}       % hyperlinks
\hypersetup{pdfstartview={XYZ null null 1.00}}
\usepackage{cleveref}
\usepackage{subfig}
\usepackage{longtable}
\usepackage{multirow}
\usepackage{multicol}
\usepackage{xcolor}
\newcommand{\cbox}[1]{\raisebox{\depth}{\fcolorbox{white}{#1}{\null}}}
\let\depth\relax

\definecolor{color0}{rgb}{0.19215686274509805, 0.50980392156862742, 0.74117647058823533}
\definecolor{color1}{rgb}{0.41960784313725491, 0.68235294117647061, 0.83921568627450982}
\definecolor{color2}{rgb}{0.61960784313725492, 0.792156862745098, 0.88235294117647056}
\definecolor{color3}{rgb} {0.77647058823529413, 0.85882352941176465, 0.93725490196078431}
\definecolor{color4}{rgb}{0.90196078431372551, 0.33333333333333331, 0.050980392156862744}
\definecolor{color5}{rgb}{0.99215686274509807, 0.55294117647058827, 0.23529411764705882}
\definecolor{color6}{rgb}{0.99215686274509807, 0.68235294117647061, 0.41960784313725491}
\definecolor{color7}{rgb}{0.99215686274509807, 0.81568627450980391, 0.63529411764705879}
\definecolor{color8}{rgb}{0.19215686274509805, 0.63921568627450975, 0.32941176470588235}
\definecolor{color9}{rgb}{0.45490196078431372, 0.7686274509803922, 0.46274509803921571}
\definecolor{color10}{rgb}{0.63137254901960782, 0.85098039215686272, 0.60784313725490191}
\definecolor{color11}{rgb}{0.7803921568627451, 0.9137254901960784, 0.75294117647058822}
\definecolor{color12}{rgb}{0.45882352941176469, 0.41960784313725491, 0.69411764705882351}
\definecolor{color13}{rgb}{0.61960784313725492, 0.60392156862745094, 0.78431372549019607}
\definecolor{color14}{rgb}{0.73725490196078436, 0.74117647058823533, 0.86274509803921573}
\definecolor{color15}{rgb}{0.85490196078431369, 0.85490196078431369, 0.92156862745098034}
\definecolor{color16}{rgb}{0.38823529411764707, 0.38823529411764707, 0.38823529411764707}
\definecolor{color17}{rgb}{0.58823529411764708, 0.58823529411764708, 0.58823529411764708}
\definecolor{color18}{rgb}{0.74117647058823533, 0.74117647058823533, 0.74117647058823533}
\definecolor{color19}{rgb}{0.85098039215686272, 0.85098039215686272, 0.8509803921568627}

\begin{document}

%\jvol{00} \jnum{00} \jyear{2014} \jmonth{October}

\title{Uncovering the mesoscale structure of the credit default swap market to improve portfolio risk modelling}

% \orcid{0000-0002-8708-2293}

\author{I. ANAGNOSTOU$^{\ast}$$\dag$$\ddag$\thanks{$^\ast$Corresponding author.
Email: i.anagnostou@uva.nl} \orcidicon{0000-0002-8708-2293} , T. SQUARTINI$\S$ \orcidicon{0000-0001-9011-966X}, D. KANDHAI$\dag$$\ddag$ \orcidicon{0000-0003-4899-7711}  and D. GARLASCHELLI$\S$$\P$ \orcidicon{0000-0001-6035-1783} \\
\affil{$\dag$Computational Science Lab, University of Amsterdam, Science Park 904, 1098 XH Amsterdam, The Netherlands\\
$\ddag$Quantitative Analytics, ING Bank, Foppingadreef 7, 1102 BD Amsterdam, The Netherlands\\
$\S$IMT School for Advanced Studies Lucca, Piazza San Francesco 19, 55100 Lucca, Italy\\
$\P$Lorentz Institute for Theoretical Physics, Leiden University, Niels Bohrweg 2, 2333 CA Leiden, The Netherlands}}

\maketitle

\begin{abstract}
{One of the most challenging aspects in the analysis and modelling of financial markets, including Credit Default Swap (CDS) markets, is the presence of an emergent, intermediate level of structure standing in between the microscopic dynamics of individual financial entities and the macroscopic dynamics of the market as a whole. This elusive, mesoscopic level of organisation is often sought for via factor models that ultimately decompose the market according to geographic regions and economic industries. However, at a more general level the presence of mesoscopic structure might be revealed in an entirely data-driven approach, looking for a modular and possibly hierarchical organisation of the empirical correlation matrix between financial time series. The crucial ingredient in such an approach is the definition of an appropriate null model for the correlation matrix. Recent research showed that community detection techniques developed for networks become intrinsically biased when applied to correlation matrices. For this reason, a method based on Random Matrix Theory has been developed, which identifies the optimal hierarchical decomposition of the system into internally correlated and mutually anti-correlated communities. Building upon this technique, here we resolve the mesoscopic structure of the CDS market and identify groups of issuers that cannot be traced back to standard industry/region taxonomies, thereby being inaccessible to standard factor models. We use this decomposition to introduce a novel default risk model that is shown to outperform more traditional alternatives.}
\end{abstract}


\begin{classcode}
C15; %Statistical Simulation Methods: General
% C38; %Classification Methods • Cluster Analysis • Principal Components • Factor Models
% C40; %Econometrics and statistical methods General
C53; %Forecasting and Prediction Methods • Simulation Methods
% C60; % Mathematical Methods • Programming Models • Mathematical and Simulation Modeling General
% C63; % Computational Techniques • Simulation Modeling
D85; %Network Formation and Analysis: Theory
G11; %Portfolio Choice • Investment Decisions
% G32; %Financing Policy • Financial Risk and Risk Management • Capital and Ownership Structure • Value of Firms • Goodwill
L14 %Transactional Relationships • Contracts and Reputation • Networks

\end{classcode}

\section{Introduction}

The financial crisis of 2007-08 laid bare the downside of a highly interconnected financial system by evidencing that each link constitutes a channel through which shocks can propagate rapidly across markets and asset classes. This has been popularised by the \emph{too-interconnected-to-fail} motto, stressing the impact that the failure of a highly central node would have on the rest of the network. Capturing financial complexity within models has been a major challenge since, for financial institutions and regulators alike.

Complexity-inspired models rest upon the evidence that economic and financial systems have many of the key properties characterising \emph{natural} complex systems: they are composed by many heterogeneous units that interact with each other in a non-linear fashion, usually in the presence of feedback \citep{amaral2004complex, mantegna2000introduction}. A tractable framework for the quantitative analysis of many complex systems is provided by \emph{networks}. Techniques from network theory have been used to study a variety of financial assets, including equities \citep{mantegna1999hierarchical,mantegna2000introduction,onnela2003dynamics}, exchange rates \citep{mcdonald2005detecting, mcdonald2008impact,fenn2012dynamical}, commodities \citep{sieczka2009correlations}, bonds \citep{bernaschi2002statistical} and interest rates \citep{di2004interest}; however, despite the growing literature on networks in finance, the effort for incorporating these techniques in models used for the pricing and risk management of individual portfolios has remained at a very early stage \citep{anagnostou2018incorporating,anagnostou2019contagious}.

In addition to the complexity of the financial system, the global financial crisis uncovered the significant weaknesses of the regulatory framework for capitalising risks from trading activities. Since the onset of the crisis, a major source of losses and of the build up of leverage transpired in the trading book: an instrumental factor was that the existing capital framework for market risk, based on the 1996 Amendment to the Capital Accord to incorporate market risks \citep{basel1996amendment}, was not able to capture some key risks. In 2009, the Basel Committee on Banking Supervision introduced a set of revisions to the market risk framework to address the most crucial shortcomings, commonly referred to as the Basel 2.5 package of reforms \citep{basel2009guidelines,basel2011revisions}. These reforms included requirements for the banks to hold additional capital against default risk and rating migration risk, known as the Incremental Risk Charge (IRC). The IRC is calculated using a value-at-risk (VaR) model at the 99.9\% confidence level over a one-year time horizon.

Although Basel 2.5 was an important improvement, some of the structural flaws of the market risk framework remained unaddressed. To this end, the Committee initiated a fundamental review of the trading book (FRTB) to enhance the design and coherence of the market risk capital standard, in line with the lessons learned from the global financial crisis \citep{basel2016minimum}. In FRTB the IRC is replaced with a Default Risk Charge (DRC) model. As an autonomous modelled approach, the IRC effectively dismisses diversification effects between credit-related risks and other risks. Moreover, the more complex IRC models were identified as source of undesired variability in the market risk weighted assets. Under the revised framework, the DRC models will measure the trading portfolio's default risk separate from all market risks. As an additional constraint, the DRC places limitations on the types of risk factors and correlations that can be used within the model. More specifically, banks must use a default simulation model with multiple systematic risk factors of two different types and default correlations must be based on credit spreads or on listed equity prices, covering a period of 10 years which includes a period of stress.

In order to meet the requirement of two different types of systematic factors for the default model, \cite{laurent2016trading} proposed to use principal components analysis (PCA) to identify the common systematic factors that drive issuer returns. While the factors obtained using this approach are explanatory in a statistical sense, they usually lack of an obvious economic meaning. Moreover, the stability of such factors over time might prove insufficient. A more common approach among practitioners is employing observable economic factors, representing the overall state of the economy or effects related to particular geographic regions or industry sectors. An example of this approach is described in \cite{wilkens2017default}.

Motivated by the absence of models encoding complexity-based `inputs', our paper takes up the challenge by focusing on one of the key aspects of complex systems, i.e. the presence of a \emph{community structure}. Detecting the presence of communities means individuating clusters of units sharing some kind of similarity: when considering financial systems, this usually boils down at identifying sets of stocks sharing similar \emph{price dynamics}. We focus on credit default swap (CDS) spread time series with the aim of individuating clusters whose similarities cannot trace back to the standard, region- and sector-wise ones. To achieve this goal, we employ a recently-proposed community detection method that take as input the empirical correlation matrix induced by a given set of time series \citep{macmahon2015community,almog2015mesoscopic} and, then, we employ the output of such a method to {identify communities of issuers and} define a novel default risk model. The variant we propose is shown to outperform more traditional versions. Our contributions in this paper are, therefore, threefold. {First, to the best of our knowledge our study is the first one that detects mesoscopic communities of issuers by applying the approach based on Random Matrix Theory (RMT) on CDS time series.
} Second, on the basis of the detected communities we derive factors and develop a {new} portfolio credit risk model that is in line with regulatory requirements for DRC calculations. Third, we set up four synthetic test portfolios and present the impact of considering different systematic factors on the quantiles of the generated loss distributions.

The remainder of the paper is organised as follows. In \Cref{sec:preliminaries} we provide a brief definition of CDS contracts and an overview of the dataset used for the present analysis. In \Cref{sec:community_detection} we describe the theoretical principles upon which the community detection method for correlation matrices employed here is based and analyse the results obtained using the data described in \Cref{sec:preliminaries}. \Cref{sec:model} is, instead, dedicated to the description of our novel default risk model whose key ingredient is represented by a previously-ignored factor, i.e. the community-driven one, as well as a simulation study on synthetic test portfolios. Finally, in \Cref{sec:conclusions} we summarise our findings and draw conclusions.

\section{Preliminary definitions}\label{sec:preliminaries}

\subsection{Credit default swaps}

A credit default swap (CDS) is a financial contract in which a protection buyer, A, buys insurance from a protection seller, B, against the default of a reference entity, C. More specifically, regular coupon payments with respect to a contractual notional and a fixed rate, the CDS spread, are swapped with a payment of  in the event of default of C, where $RR$, known as the recovery rate, is a contract parameter representing the fraction of investment which is assumed be recovered in the event of default of C.

CDS spreads reflect the market participants' view on probability of default. Thus, practitioners often rely on them to obtain market-implied parameters which are key inputs to their models. For instance, in the case of credit valuation adjustment (CVA)\footnote[1]{CVA is the difference between the risk-free portfolio value and the actual portfolio market value that takes into account the risk of a counterparty’s default. Its magnitude depends on the probability of default of the counterparty, the future exposures of the underlying
derivative or portfolio, and the loss given default.}, the default probabilities are obtained from CDS spreads. Apart from derivatives valuation, CDS spreads are used extensively for estimating correlations in market risk and capital models \citep{basel2016minimum}.

\subsection{Description of the data-set}

The raw CDS data-set is provided by Markit and consists of daily CDS spreads for a range of maturities covering the period between 1 January 2007 and 31 December 2016. Markit maintains a network of market makers who contribute quotes from their official books and records for thousands of entities on a daily basis. Using the contributed quotes, the daily CDS spreads for each entity, as well as the daily recovery rates used to price the contracts, are calculated. In addition, the data-set contains information on the names of the underlying reference entities, recovery rates, seniority of the debt on which the contract is priced on, restructuring type, number of quote contributors, region, sector, average of the ratings from Standard \& Poor’s, Moody’s, Fitch Group of each entity and currency of the quote.

Since Markit data are characterised by a number of attributes, it is possible to have multiple CDS data series for the same issuer. In order to obtain unique, representative time series, we apply a set of selection criteria. First, we select the CDS spreads of entities for the five-year tenor for our analysis; it is observed that Markit's raw data is more complete for this tenor, since five-year CDS contracts are the most liquid. For the same reason, we select senior unsecured debt for corporates and foreign debt for sovereigns. Finally, we set up a hierarchy for the document clause/restructuring type which defines what constitutes a credit event and select the series denominated in Euro for the European issuers and in U.S. dollar for issuers from the rest of the world. Besides the above steps, we apply a couple of additional filtering steps to the CDS data to retain the most liquid quotes. After applying these pre-processing steps, we are left with a total of 786 entities and 2608 trading days. The distribution of issuers across regions and sectors is shown in \Cref{tab:data_distribution}.
\begin{table}
\begin{center}
\begin{minipage}{\textwidth}
\tbl{Distribution of 786 entities across regions and sectors.}
{\begin{tabular}{@{}llr}\toprule
Region & Sector & $N$ \\
\colrule
Africa     & Consumer Services           & 1  \\
           & Government                  & 4  \\
Asia       & Basic Materials             & 6  \\
           & Consumer Goods              & 20 \\
           & Consumer Services           & 9  \\
           & Energy                      & 8  \\
           & Financials                  & 29 \\
           & Government                  & 14 \\
           & Industrials                 & 15 \\
           & Technology                  & 7  \\
           & Telecommunications Services & 9  \\
           & Utilities                   & 15 \\
Eastern Europe       & Financials                  & 3  \\
           & Government                  & 11 \\
Europe     & Basic Materials             & 17 \\
           & Consumer Goods              & 29 \\
           & Consumer Services           & 33 \\
           & Energy                      & 7  \\
           & Financials                  & 63 \\
           & Government                  & 13 \\
           & Health Care                  & 5  \\
           & Industrials                 & 26 \\
           & Technology                  & 6  \\
           & Telecommunications Services & 15 \\
           & Utilities                   & 25 \\
India      & Basic Materials             & 1  \\
           & Financials                  & 4  \\
           & Government                  & 1  \\
           & Industrials                 & 1  \\
Latin America   & Consumer Services           & 1  \\
           & Energy                      & 1  \\
           & Government                  & 11 \\
Middle East & Government                  & 6  \\
           & Health Care                  & 1  \\
           & Utilities                   & 2  \\
North America     & Basic Materials             & 25 \\
           & Consumer Goods              & 53 \\
           & Consumer Services           & 46 \\
           & Energy                      & 33 \\
           & Financials                  & 56 \\
           & Government                  & 2  \\
           & Health Care                  & 27 \\
           & Industrials                 & 40 \\
           & Technology                  & 17 \\
           & Telecommunications Services & 16 \\
           & Utilities                   & 27 \\
Oceania    & Basic Materials             & 2  \\
           & Consumer Goods              & 2  \\
           & Consumer Services           & 4  \\
           & Energy                      & 2  \\
           & Financials                  & 9  \\
           & Government                  & 1  \\
           & Industrials                 & 3  \\
           & Telecommunications Services & 2 \\
\botrule
\end{tabular}}
\label{tab:data_distribution}
\end{minipage}
\end{center}
\end{table}

\section{Community detection {on CDS} correlation matrices}\label{sec:community_detection}

\subsection{Methods}\label{subsec:community_detection_methods}
\paragraph{Basic notation} In this subsection we introduce the basic notation and describe the community-detection method for time series introduced in \cite{macmahon2015community}. {Financial} markets are represented as a set of time series $X_1\dots X_N$, each one encoding the temporally ordered activity of the $i$-th unit of the system over, say, $T$ time-steps, i.e.

\begin{equation}
X_i\equiv\{x_i(1)\dots x_i(T)\},\:\forall\:i.
\end{equation}

In our case $i$ is a {credit} issuer. The mutual interactions between the series considered above are summed up by an $X\times N$ \emph{correlation matrix}, i.e. a matrix $\mathbf{C}$ whose generic entry $C_{ij}$ reads

\begin{equation}
C_{ij}=\frac{\text{Cov}[X_i,X_j]}{\sqrt{\text{Var}[X_i]\cdot\text{Var}[X_j]}},\:\forall\:i,j
\end{equation}
i.e. the Pearson coefficient between series $i$ and $j$, with

\begin{equation}
\text{Cov}[X_i,X_j]=\overline{X_i\cdot X_j}-\overline{X_i}\cdot\overline{X_j},\:\forall\:i,j
\end{equation}
and 
\begin{equation}
\text{Var}[X_i]=\overline{X_i^2}-\overline{X_i}^2,\:\forall\:i;
\end{equation}
in the above equations, the bar is assumed to denote a temporal average, i.e.

\begin{eqnarray}
\overline{X_i}&=&\frac{\sum_{t=1}^Tx_i(t)}{T},\:\forall\:i,\\
\overline{X_i^2}&=&\frac{\sum_{t=1}^Tx_i^2(t)}{T},\:\forall\:i,\\
\overline{X_i\cdot X_j}&=&\frac{\sum_{t=1}^Tx_i(t)\cdot x_j(t)}{T},\:\forall\:i,j.
\end{eqnarray}

As frequently done in order to filter out the inherent heterogeneity of time series, each series $X_i$ has been \emph{standardised} by subtracting the temporal average $\overline{X_i}$ and dividing the result by the standard deviation $\sigma_i=\sqrt{\text{Var}[X_i]}$; in other words, $X_i$ has been redefined as $(X_i-\overline{X_i})/\sigma_i$ in a such a way to ensure that $\overline{X_i}=0$, $\text{Var}[X_i]=1$ and $C_{ij}=\text{Cov}[X_i,X_j]=\overline{X_i\cdot X_j}$.

One of most challenging problems in the field of complex systems is that of extracting information from the matrix $\mathbf{C}$. As we are considering financial markets, we would be interested in individuating sets of issuers sharing a similar CDS spread dynamics. More formally, this amounts at inspecting the community structure of the considered set of issuers, i.e. the presence of (internally cohesive) \emph{modules of issuers}.

Over the past years, several techniques to retrieve information regarding the modularity of multiple time series have been proposed (see \cite{macmahon2015community} and references therein). One of the most promising approach is that of applying community detection techniques to empirical correlation matrices: however, as existing methods are tailored on graphs, they suffer from statistical bias whenever applied `as they are' to correlation matrices (see \cite{macmahon2015community} and references therein).

Recently, a method based on Random Matrix Theory (RMT) \citep{potters2005financial,mehta2004random} has been proposed \citep{macmahon2015community}: when applied to financial time series, this algorithm has been proven to be able to capture the dynamical modularity of real markets, by identifying clusters of stocks which are positively correlated \emph{internally} but anti-correlated with each other. In what follows, we will provide a brief explanation of {the method. More details can be found in the original reference \citep{macmahon2015community}}.

\paragraph{Spectral analysis of random correlation matrices} Let us start by inspecting the properties of random correlation matrices. The latter are constructed by considering $N$ completely random time series of length $T$ (more precisely, time series whose entries are independent, identically distributed random variables with zero mean and finite variance): the matrix encoding the correlations of this set of series is a $N\times N$ Wishart matrix, whose eigenvalues follow (in the limits $N \to +\infty$ and
$T \to +\infty$ with $1<T/N<+\infty$) the so-called Marcenko-Pastur distribution \citep{laloux1999noise}, reading

\begin{equation}\label{eq:marchenko}
\rho(\lambda)=\frac{T}{N}\frac{\sqrt{(\lambda_+-\lambda)(\lambda-\lambda_-)}}{2\pi\lambda}\quad\text{if}\quad\lambda_-\leq\lambda\leq\lambda_+
\end{equation}
and $\rho(\lambda)=0$ otherwise, with $\lambda_+$ and $\lambda_-$ being, respectively, the maximum and the minimum eigenvalue:

\begin{equation}
\lambda_{\pm}=\left[1\pm\sqrt{\frac{N}{T}}\right]^2.
\end{equation}

\begin{figure}
\centering
\includegraphics[width=0.49\textwidth]{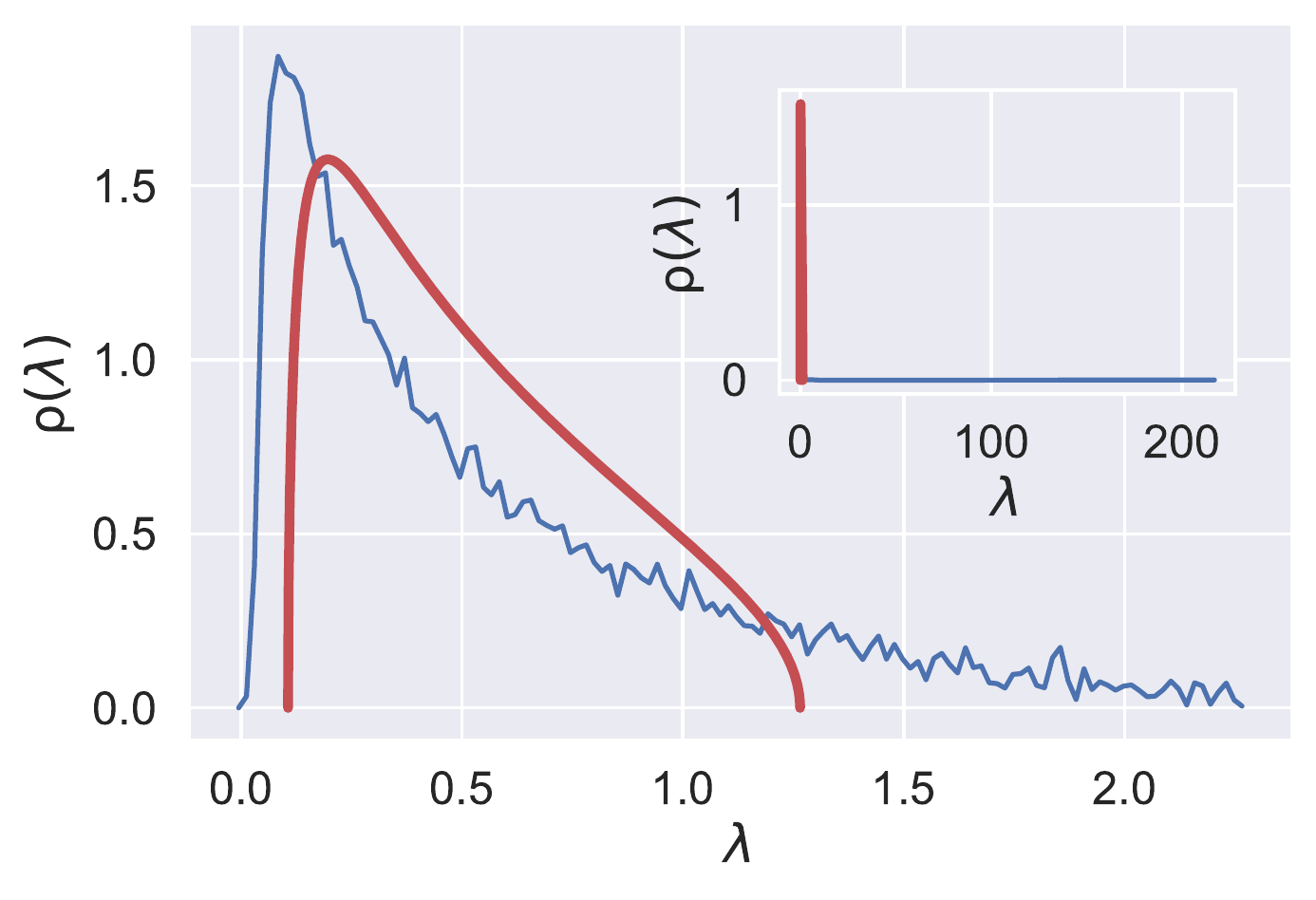}
\includegraphics[width=0.49\textwidth]{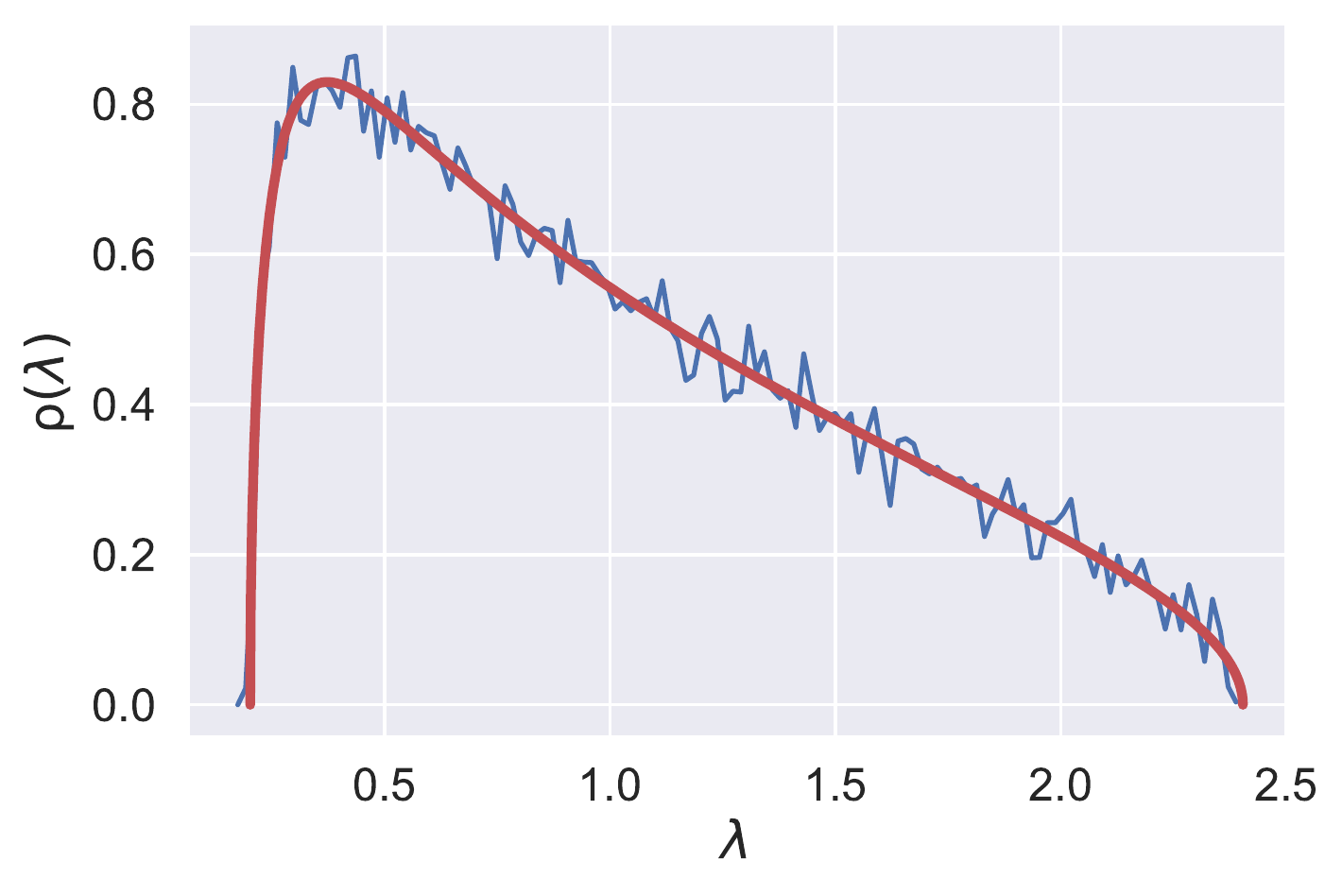}
\caption{Left panel: smoothed eigenvalue density of the correlation matrix extracted from the CDS spreads of $N=786$ issuers during the period 2007-2016 ($T=2608$); for comparison we have plotted the Marcenko-Pastur density function \labelcref{eq:marchenko} coming from $N$ uncorrelated series of duration $T$. Inset: same plot, but including the highest eigenvalue corresponding to the `market'. Right panel: eigenvalues density distribution of randomised CDS data. The figure shows the agreement between empirical (in blue) and random (in red) distributions once the original data are shuffled.}
\label{fig:eigenvalues}
\end{figure}

The method we are going to describe implements the idea that, while the eigenvalues of an empirical correlation matrix falling within these boundaries can be attributed to random noise, any eigenvalue smaller than $\lambda_-$ and larger than $\lambda_+$ is to be considered as representing some meaningful structure in the data. The result above leads to the possibility of expressing any empirical correlation matrix as the sum of two components, i.e.

\begin{equation}
\mathbf{C}=\mathbf{C}^{(r)}+\mathbf{C}^{(s)} 
\end{equation}
where $\mathbf{C}^{(r)}$ is the tensor \emph{random component}, induced by the eigenvalues in the random bulk (i.e. $\forall\:i$ such that $\lambda_i\in[\lambda_-,\lambda_+]$) and reading

\begin{equation}
\mathbf{C}^{(r)}\equiv\sum_{\substack{i\\(\lambda_-\leq\lambda_i\leq\lambda_+)}}\lambda_i|v_i\rangle\langle v_i|
\end{equation}
and $\mathbf{C}^{(s)}$ is the tensor \emph{structural component}, aggregated from the remaining eigenvalues. The structural component $\mathbf{C}^{(s)}$ can be further subdivided, upon considering the existence of the so called \emph{market mode}. As it has been shown in a number of previous studies \citep{macmahon2015community,almog2015mesoscopic,laloux1999noise}, a representative characteristic of the spectrum of empirical correlation matrices is the existence of an eigenvalue $\lambda_m$ which is orders of magnitude larger that the remaining ones; in case financial stocks are considered, such a leading eigenvalue embodies the common factor driving all the constituents of a given market. As the effect of $\lambda_m$ is that of pushing all nodes into the same community, we need to properly discount it, by focusing on the `reduced' portion of the structured spectrum defined by the condition $\lambda_i\in(\lambda_+,\lambda_m)$. This leads to a further decomposition of the correlation matrix, i.e.

\begin{equation}\label{eq:superposition}
\mathbf{C}=\mathbf{C}^{(r)}+\mathbf{C}^{(g)}+\mathbf{C}^{(m)}
\end{equation}
where
\begin{equation}
\mathbf{C}^{(m)}\equiv\lambda_m|v_m\rangle\langle v_m|
\end{equation}
represents the tensor portion induced by the market mode and
\begin{equation}
\mathbf{C}^{(g)}\equiv\sum_{\substack{i\\(\lambda_+<\lambda_i<\lambda_m)}}\lambda_i|v_i\rangle \langle v_i|
\end{equation}
represents the tensor portion filtered from both the random noise and the common factor. As a consequence, the correlations encoded in $\mathbf{C}^{(g)}$ are neither at the individual level nor at the level of the entire market but at the level of \emph{groups of stocks} (i.e. at the \emph{mesoscale} in the network jargon). Remarkably, the eigenvectors contributing to $\mathbf{C}^{(g)}$ have alternating signs, allowing for the detection of groups affected in a similar manner by some (other) common factors \citep{macmahon2015community,potters2005financial}.

In Figure \ref{fig:eigenvalues} we plot the eigenvalue density distribution characterising the empirical correlation matrix of $N=786$ CDS spreads (corresponding to $T=2608$ daily log-returns for the period 2007-2016) together with the Marcenko-Pastur distribution coming from $N$ totally uncorrelated series of duration $T$ (shown in red). The maximum expected eigenvalue amounts at approximately $\lambda_+=1.27$. The inset is the fully zoomed-out version of the plot, illustrating that the empirical correlation matrix has a maximum eigenvalue of about $\lambda_m=216$ (i.e. the market mode), in addition to a few other eigenvalues lying between $\lambda_+$ and $\lambda_m$. The eigenvector $|v_m\rangle$ corresponding to $\lambda_m$ has all positive signs. We also inspect whether the system follows the Marcenko-Pastur distribution once the original data re shuffled. To this aim, we, first, randomly permute the entries of each time series separately, thus destroying the daily correlations; then, we check if the eigenvalues of the correlation matrix of the shuffled set of series follow the Marcenko-Pastur distribution: as fig. \ref{fig:eigenvalues} shows, this is indeed the case.

\paragraph{Community detection on filtered correlation matrices} Community detection is an active field of research within network theory. Among the many proposed approaches, the most popular one is based on the maximisation of the quantity known as \emph{modularity}, a score function measuring the optimality of a given partition by comparing the empirical pattern of interconnections with the one predicted by a properly-defined benchmark model. It is defined as

\begin{equation}
Q(\bm{\sigma})=\frac{1}{||A||}\sum_{i,j=1}^N\left[a_{ij}-\langle a_{ij}\rangle\right]\delta(\sigma_i,\sigma_j)
\label{eq:modularity}
\end{equation}
where $a_{ij}$ is the generic entry of the network adjacency matrix $\mathbf{A}$, $\langle a_{ij}\rangle$ is the probability that nodes $i$ and $j$ establish a connection according to the chosen benchmark (i.e. the expectation of whether a link exists or not under some suitable null hypothesis), $\bm{\sigma}$ is the $N$-dimensional vector encoding the information carried by a given partition (the $i$-th component, $\sigma_i$, denotes the module to which node $i$ is assigned) and the Kronecker delta $\delta(\sigma_i,\sigma_j)$ ensures that only the nodes within the same modules provide a positive contribution to the sum. The normalisation factor $||A||=\sum_{i,j=1}^Na_{ij}$ guarantees that $-1\leq Q(\bm{\sigma})\leq1$ (when undirected networks are considered, it equals twice the number of connections).

Much of the previous research in community detection for financial time series has explored the approach of considering the empirical correlation matrix $\mathbf{C}$ as a weighted network and searching for communities by using the weighted extension of the modularity, defined by posing $a_{ij}\equiv C_{ij}$ and 
\begin{equation}
\langle a_{ij}\rangle=\langle C_{ij}\rangle=\frac{s_is_j}{2W},\:\forall\:i,j
\end{equation}
with $s_i=\sum_{l=1}^NC_{il}=\text{Cov}[X_i,X_{tot}]$ and $2W=\sum_{i,j=1}^NC_{ij}=\text{Var}[X_{tot}]$ (where $X_{tot} = \lbrace x_{tot}(1)\dots x_{tot}(T)\rbrace$ is the time series of the total increment $x_{tot}(t)\equiv\sum_{j=1}^Nx_j(t)$). According to \cite{macmahon2015community}, however, this approach may lead to biased results. As a consequence, a null model encoding the spectral properties of correlation matrices has been proposed, i.e.

\begin{equation}
\langle a_{ij}\rangle=\langle C_{ij}\rangle=C_{ij}^{(r)}+C_{ij}^{(m)},\:\forall\:i,j
\end{equation}
in turn leading to the following redefinition of modularity
\begin{equation}
Q(\bm{\sigma})=\frac{1}{||C||}\displaystyle\sum_{i,j=1}^N\left[C_{ij}-\left(C_{ij}^{(r)}+C_{ij}^{(m)}\right)\right]\delta(\sigma_i,\sigma_j)
\label{eq:new_modularity}
\end{equation}
where $||C||=\sum_{i,j=1}^NC_{ij}=\text{Var}[X_{tot}]$ and $X_{tot} = \lbrace x_{tot}(1)\dots x_{tot}(T)\rbrace$ is the time series of the total increment $x_{tot}(t)\equiv\sum_{j=1}^Nx_j(t)$. The maximisation of this `novel' quantity outputs a partition of a given set of time series upon filtering out the random noise and the market component. It is worth noting that the selection of $||C||$ in \Cref{eq:new_modularity} automatically controls for the volatility of the system, a particularly desirable feature when analysing the evolution of the community structure of highly fluctuating systems. 

\subsection{Results}\label{sec:results}

\begin{figure}
\centering
\hspace*{-0.05cm}
\begin{minipage}{.45\linewidth}
\centering
\subfloat[]{\label{communities:a}\includegraphics[width=\textwidth]{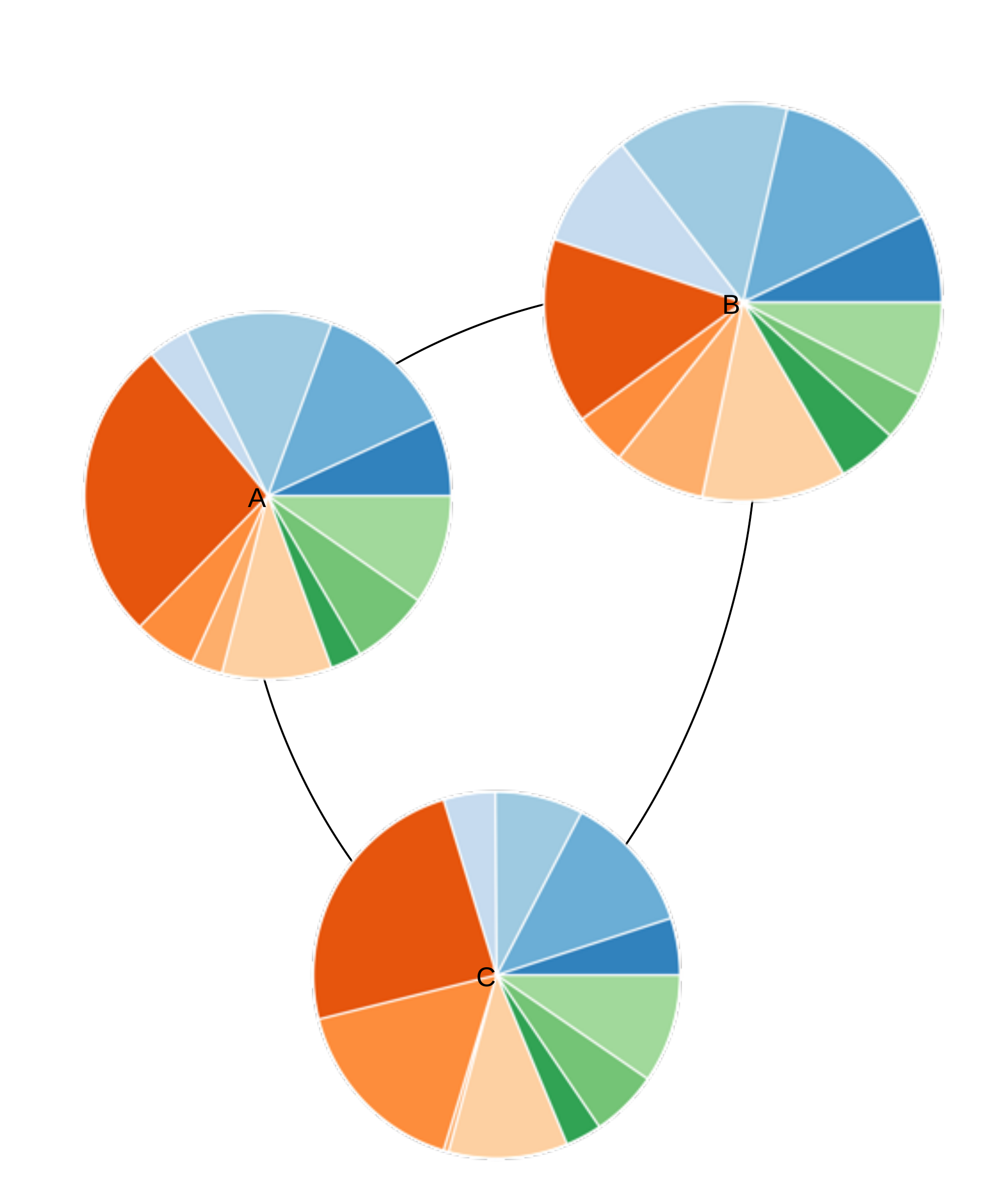}}
\end{minipage}
\begin{minipage}{.45\linewidth}
\centering
\hspace{0.05cm}
\vspace*{0.63cm}
  \subfloat[]{\label{communities:b}\includegraphics[width=\textwidth]{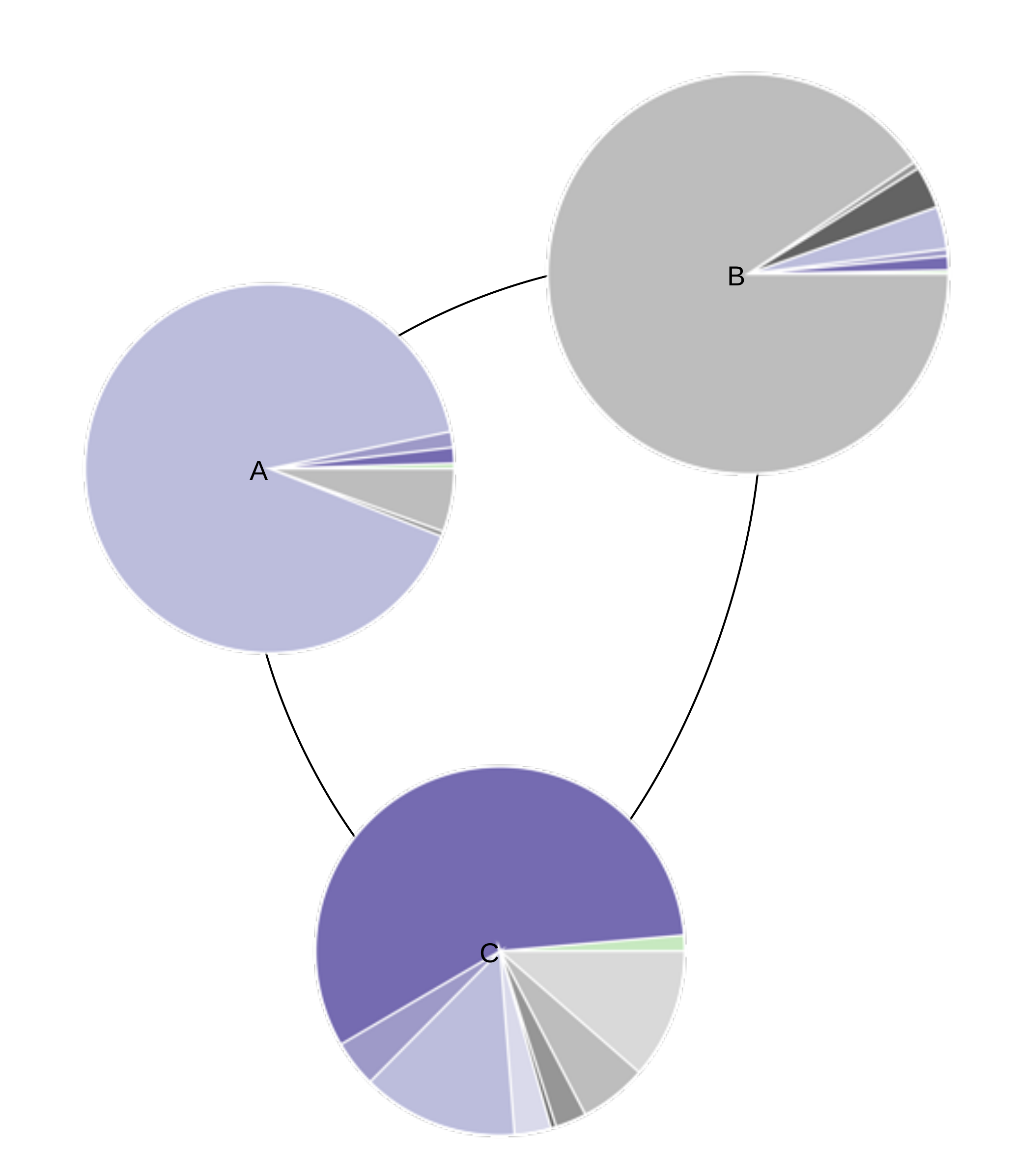}}
\end{minipage}
\caption{CDS market community structure. The figure shows the communities detected on CDS spread data for 786 issuers during the period 1 January 2007-31 December 2016. The communities were generated by the modified Louvain algorithm \citep{macmahon2015community} using daily log-returns. Individual communities are labelled A, B, C and the pie charts represent the relative composition of each community with respect to the sector (left panel) and region (right panel) of the component entities.}
\label{fig:communities}
\end{figure}

\paragraph{Community detection} We now proceed with the application of the methodology described in \Cref{subsec:community_detection_methods}. \Cref{fig:communities} shows the output of the Louvain algorithm when applied to the daily CDS spread data of 786 issuers, covering the period between 1 January 2007 and 31 December 2016. The algorithm is making use of the modularity defined in \Cref{eq:new_modularity}, which is able to discount random as well as market-wide effects. With the CDS data, we obtain three mesoscopic communities, labelled A, B and C, characterised (as explained in the previous sections) by \emph{positive correlations within them} and \emph{negative correlations between them}. The pie charts represent the composition of each community according to the industry and region of the constituent issuers for \Cref{communities:a} and \Cref{communities:b} respectively. The colour legends can be found in \Cref{tab:sector_colours,tab:region_colours}.

From the data in \Cref{communities:a}, it is apparent that every community contains a range of issuers from all industry sectors: no pattern of association between sector and community structure is immediately evident. For Community A, it can be seen that over a quarter of issuers are classified as Financials (\cbox{color4}), while another quarter of the community comprises issuers from the Consumer Goods (\cbox{color1}) and Consumer Services (\cbox{color2}) sectors; the rest of the industry sectors are represented with lower percentages, ranging from slightly less than 10\% for Utilities (\cbox{color10}) to approximately 3\% for Technology (\cbox{color8}). Issuers from Financials (\cbox{color4}) are slightly less frequent in Community B, accounting for approximately 15\% of the total issuers, while consumer Goods (\cbox{color1}), Consumer Services (\cbox{color2}), and Industrials (\cbox{color7}) sectors follow with similar percentages; the rest of the community consists of issuers from all sectors with lower percentages, such as Energy (\cbox{color3}) and Utilities (\cbox{color10}) with slightly less than 10\% each. Almost 40\% of Community C consists of  Financials (\cbox{color4}) and Government (\cbox{color5}) issuers: interestingly, the proportion of Government (\cbox{color5}) issuers in Community C is significantly higher than the corresponding proportion in Communities A and B and almost twice as high as the one in the full sample of 786 issuers; issuers from the Consumer Goods (\cbox{color1}) and Industrials (\cbox{color7}) sectors are also quite frequent, constituting approximately a quarter of Community C.

\begin{table}
\begin{center}
\begin{minipage}{\textwidth}
\tbl{The 11 industry sectors with the colour representation used to highlight the sectors in the following figures.}
{\begin{tabular}{@{}llll}\toprule
Region & $N$ & Sector & $N$ \\
\colrule
 Basic Materials: &\cbox{color0}& Health Care: &\cbox{color6} \\
 Consumer Goods: & \cbox{color1}& Industrials: &\cbox{color7}   \\
 Consumer Services: & \cbox{color2}& Technology: &\cbox{color8}   \\
 Energy:   & \cbox{color3}&Telecommunications Services &\cbox{color9}  \\
 Financials: & \cbox{color4}& Utilities: &\cbox{color10}    \\
 Government &\cbox{color5}&  &    \\
\botrule
\end{tabular}}
\label{tab:sector_colours}
\end{minipage}
\end{center}
\end{table}

% \begin{table}[t!]
% \caption{The 11 industry sectors with the colour representation used to highlight the sectors in the following figures.}
% \label{tab:sector_colours}
% \centering
% \begin{tabular}{llll}
% \hline\\
%   Basic materials: &\cbox{color0}& Health care: &\cbox{color6} \\
%  Consumer goods: & \cbox{color1}& Industrials: &\cbox{color7}   \\
%  Consumer services: & \cbox{color2}& Technology: &\cbox{color8}   \\
%  Energy:   & \cbox{color3}&Telecommunication services &\cbox{color9}  \\
%  Financials: & \cbox{color4}& Utilities: &\cbox{color10}    \\
%  Government &\cbox{color5}&  &    \\
%  \\
%  \hline
% \end{tabular}
% \end{table}

\begin{table}
\begin{center}
\begin{minipage}{\textwidth}
\tbl{The 9 regions with the colour representation used to highlight the sectors in the following figures.}
{\begin{tabular}{@{}llll}\toprule
Region & $N$ & Sector & $N$ \\
\colrule
  Africa: &\cbox{color11}& Latin America: &\cbox{color16} \\
 Asia: & \cbox{color12}& Middle East: &\cbox{color17}   \\
 Eastern Europe: & \cbox{color13}& North America: &\cbox{color18}   \\
 Europe:   & \cbox{color14}&Oceania &\cbox{color19} \\
 India: & \cbox{color15}& &  \\
\botrule
\end{tabular}}
\label{tab:region_colours}
\end{minipage}
\end{center}
\end{table}

% \begin{table}[t!]
% \caption{The 9 regions with the colour representation used to highlight the sectors in the following figures.}
% \label{tab:region_colours}
% \centering
% \begin{tabular}{llll}
% \hline\\
%   Africa: &\cbox{color11}& Latin America: &\cbox{color16} \\
%  Asia: & \cbox{color12}& Middle East: &\cbox{color17}   \\
%  Eastern Europe: & \cbox{color13}& North America: &\cbox{color18}   \\
%  Europe:   & \cbox{color14}&Oceania &\cbox{color19} \\
%  India: & \cbox{color15}& &  \\
%  \\
% \hline
% \end{tabular}
% \end{table}

We now turn to the relative breakdown of the issuers of each community according to region. As \Cref{communities:b} demonstrates, the identified communities display a high degree of overlap with region classification. Communities A and B are dominated by the regions Europe (\cbox{color14}) and North America (\cbox{color18}) respectively. On the other hand, Community C contains the bulk of issuers from Asia (\cbox{color12}), while issuers from Oceania (\cbox{color19}) and India (\cbox{color15}) are represented exclusively in this community. It is interesting, however, that a considerable amount of issuers from Europe (\cbox{color14}) and North America (\cbox{color18}) can be found in Community C, meaning that over the course of the ten-year period under analysis they were more correlated with issuers from Asia (\cbox{color12}) or Oceania (\cbox{color19}) than with most of the issuers located in their own region. A possible explanation is that some issuers such as international banks registered in Europe and North America have considerable exposure to Asia-Pacific.

\paragraph{Hierarchical community structure} The methodology described in \Cref{subsec:community_detection_methods} can be applied iteratively, to individuate smaller subcommunities which may be nested inside larger communities. As the leading eigenvalue of the correlation matrix represents something that all issuers in the market have in common, the leading eigenvalue of the correlation matrix restricted to a specific individual community can be interpreted as something that all issuers in \emph{that} community have in common: by washing out the effects of this `community mode', one can detect the `residual' modular structure (internal to each community). Naturally, subcommunities within each parent community are anticorrelated with each other while being positively correlated internally.

\begin{figure}
\centering
\includegraphics[width=\textwidth]{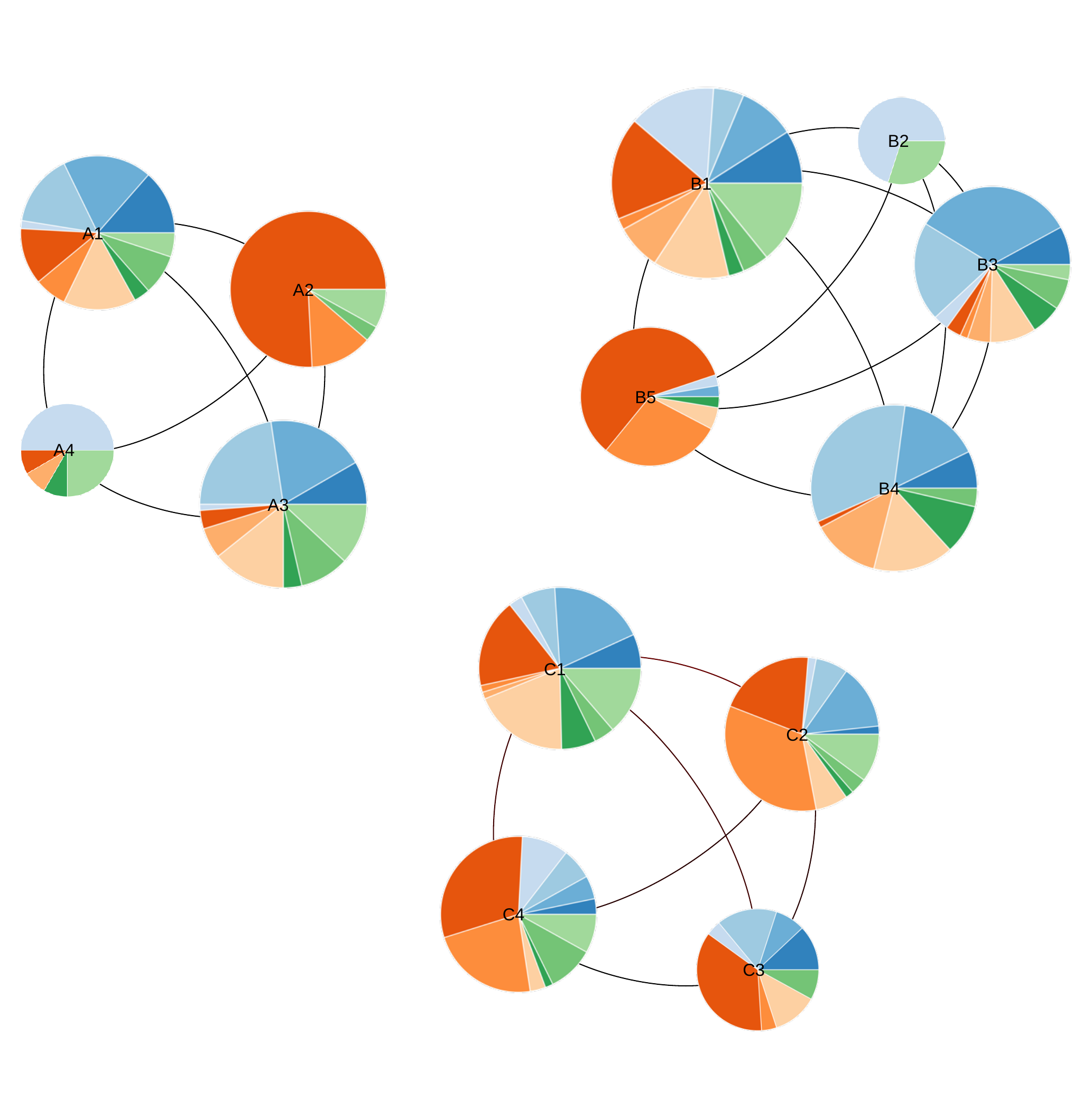}
\caption{Subcommunity structure of the three communities of the CDS market by sector. Although at first glance the subcommunities seem quite heterogeneous with respect to sector, after close inspection it can be seen that some of them are dominated by certain sectors, for example A2 and B5 are dominated by Financials (\cbox{color4}) and Government (\cbox{color5}), while A4 and B2 are dominated by Energy (\cbox{color3}) and Utilities (\cbox{color10}).
}\label{fig:subcom_sec}
\end{figure}

\begin{figure}
\centering
\includegraphics[width=\textwidth]{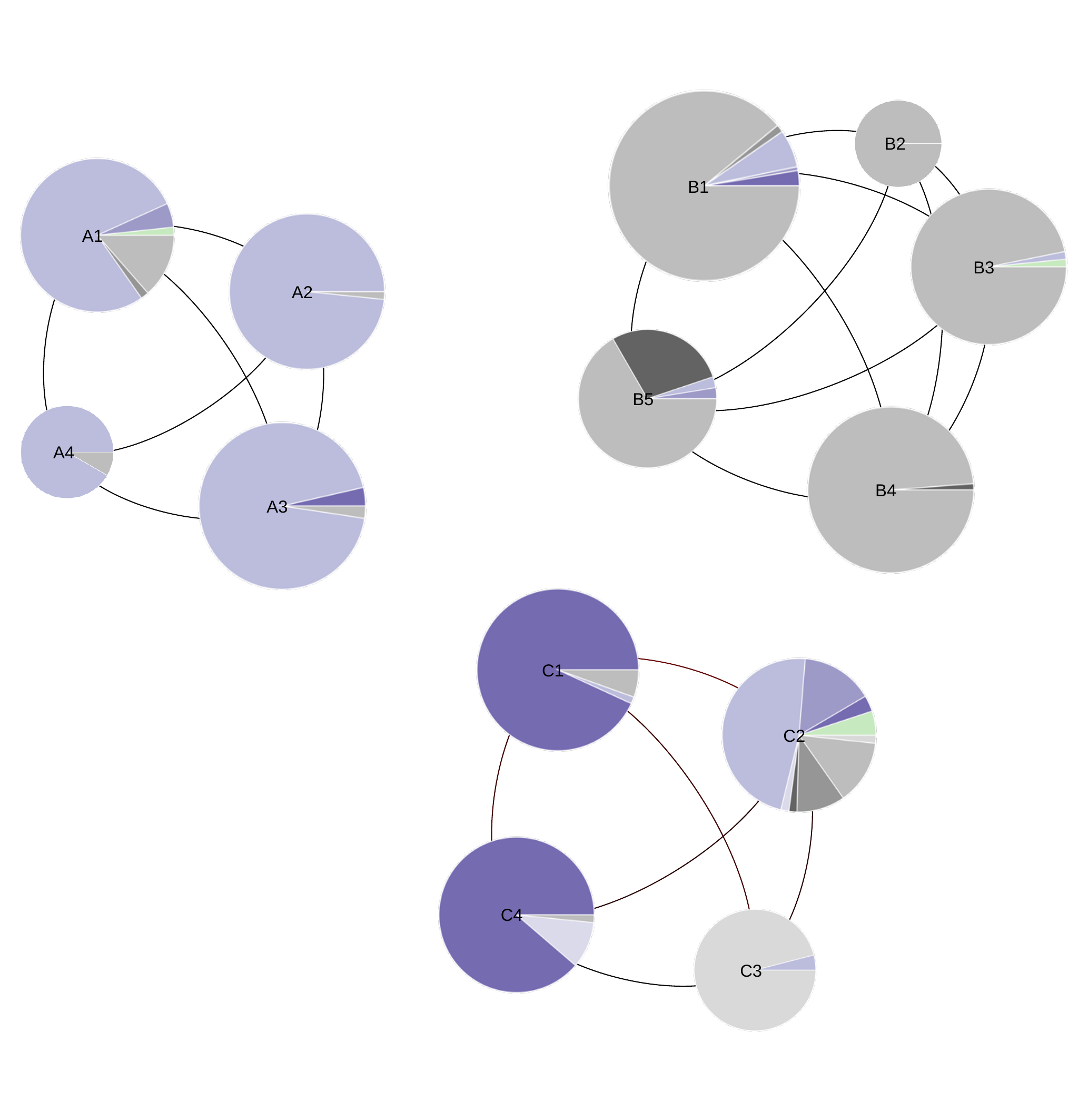}
\caption{Subcommunity structure of the three communities of the CDS market by region. The overlap between region and subcommunities is better than the one between sectors and subcommunities.}\label{fig:subcom_reg}
\end{figure}

The results of a single iteration over the communities A, B, and C (summarised in \Cref{fig:communities}) are illustrated in \Cref{fig:subcom_sec,fig:subcom_reg}. It can be seen from the graph that the degree of overlap with the sector-based classification is higher for subcommunities than for communities. Moreover, the overlap with the region-based classification is even more pronounced.

Community A is divided in four subcommunities labelled A1 through A4. A1 and A3 contain a range of issuers from all sectors, while A2 is dominated by Financials (\cbox{color4}) and three quarters of A4 is constituted by Energy (\cbox{color3}) and Utilities (\cbox{color10}). In all four subcommunities, Europe (\cbox{color14}) is the leading region and there is a small percentage of issuers from North America (\cbox{color18}); moreover, Middle East (\cbox{color17}), Eastern Europe (\cbox{color13}) and Africa (\cbox{color11}) are represented exclusively in A1 while issuers from Asia (\cbox{color12}) can be found  exclusively in A3. 

Community B is split in five subcommunities. B1 is quite heterogeneous, including issuers from all different industry sectors. B2 includes issuers from the Energy (\cbox{color3}) and Utilities (\cbox{color10}) sectors. Approximately half of B3 and B4 include are constituted by Consumer Goods (\cbox{color1}) and Consumer Services (\cbox{color2}) issuers, with Industrials (\cbox{color7}) and Health Care (\cbox{color6}) accounting for another quarter of B4. Finally, B5 is dominated by Financials (\cbox{color4}) and Government (\cbox{color5}). From a geographic perspective, subcommunities of B are all dominated by North America (\cbox{color18}), with some issuers from Europe (\cbox{color14}) being present in B1 and to a lesser extent in B3 and B5. B2 contains issuers only from North America (\cbox{color18}) and, with very few exceptions, the same holds for B3 and B4 as well. Most of the issuers from Latin America (\cbox{color16}) can be found in B5 with almost a third of this subcommunity being from that region.

Moving to the four subcommunities of Community C, almost half of the issuers in subcommunity C1 are equally split among Consumer Goods (\cbox{color1}) and Industrials (\cbox{color7}). Issuers from the Government (\cbox{color5}) and Financials (\cbox{color4}) sectors constitute more than half of subcommunities C2 and C4. Financials (\cbox{color4}) are frequent in subcommunity C3 as well, followed by Consumer Services (\cbox{color2}), i.e. the second most frequent industry sector. In terms of regional classification, although Community C is more heterogeneous than A and B, its subcommunities reveal that issuers from Asia (\cbox{color12} are concentrated in subcommunities C1 and C4, with C4 containing almost all the issuers from India (\cbox{color15}). C3 consists of issuers from Oceania (\cbox{color19}), with a small number of issuers from Europe (\cbox{color14}). The bulk of the European issuers in community C is concentrated in C2, making up almost 50\% of the subcommunity. The rest of C2 consists of issuers from Eastern Europe (\cbox{color13}) and North America (\cbox{color18}) with about 15\% each, and Middle East (\cbox{color17}) with slightly over 10\%.

In summary, these results provide important insights into the structure of the CDS market. It is suggested that the three communities initially detected are quite heterogeneous as far as industry sectors are concerned, while they overlap to a greater extent when considering a regional-based classification of issuers. Interestingly, although Communities A and B are dominated by Europe (\cbox{color14}) and North America (\cbox{color18}), some European and North American issuers are clustered with issuers from Asia (\cbox{color12}) and Oceania (\cbox{color19}) in Community C. At the second iteration of the method, some of the obtained subcommunities are dominated by certain sectors: for instance, A2 and B5 are dominated by Financials (\cbox{color4}) and Government (\cbox{color5}), while A4 and B2 are dominated by Energy (\cbox{color3}) and Utilities (\cbox{color10}). The overlap between regions and subcommunities is even better than the one between sectors and subcommunities. These results have implications for the management of portfolios of credit risky instruments, demonstrating that after global effects have been filtered out, default risk dependence is related to regional effects to a larger extent than to sectoral effects. This seems to be in line with previous evidence from equity markets \citep{heston1994does}, indicating that industry-specific effects are less significant than region effects. In addition, as far as default risk is concerned, it appears that neither diversification over regions alone nor diversification over industries alone can achieve the optimal diversification benefits, a result that is aligned with \cite{aretz2013common}.

\begin{figure}
\centering
\begin{minipage}{.49\linewidth}
\vspace*{0.5cm}
\centering
\subfloat[]{\label{resolution:a}\includegraphics[width=\textwidth]{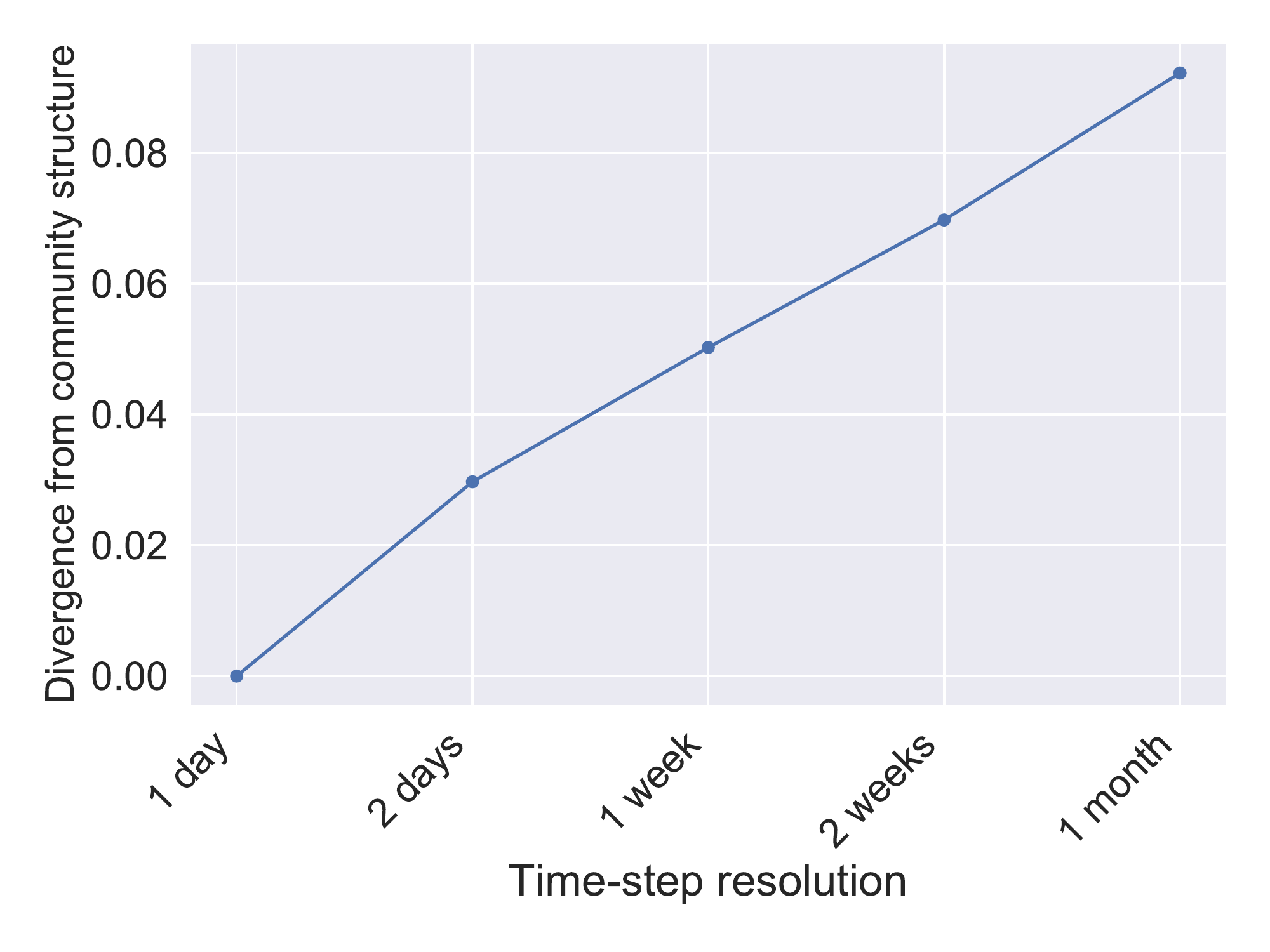}}
\end{minipage}
\begin{minipage}{.50\linewidth}
\centering
\subfloat[]{\label{resolution:b}\includegraphics[width=\textwidth]{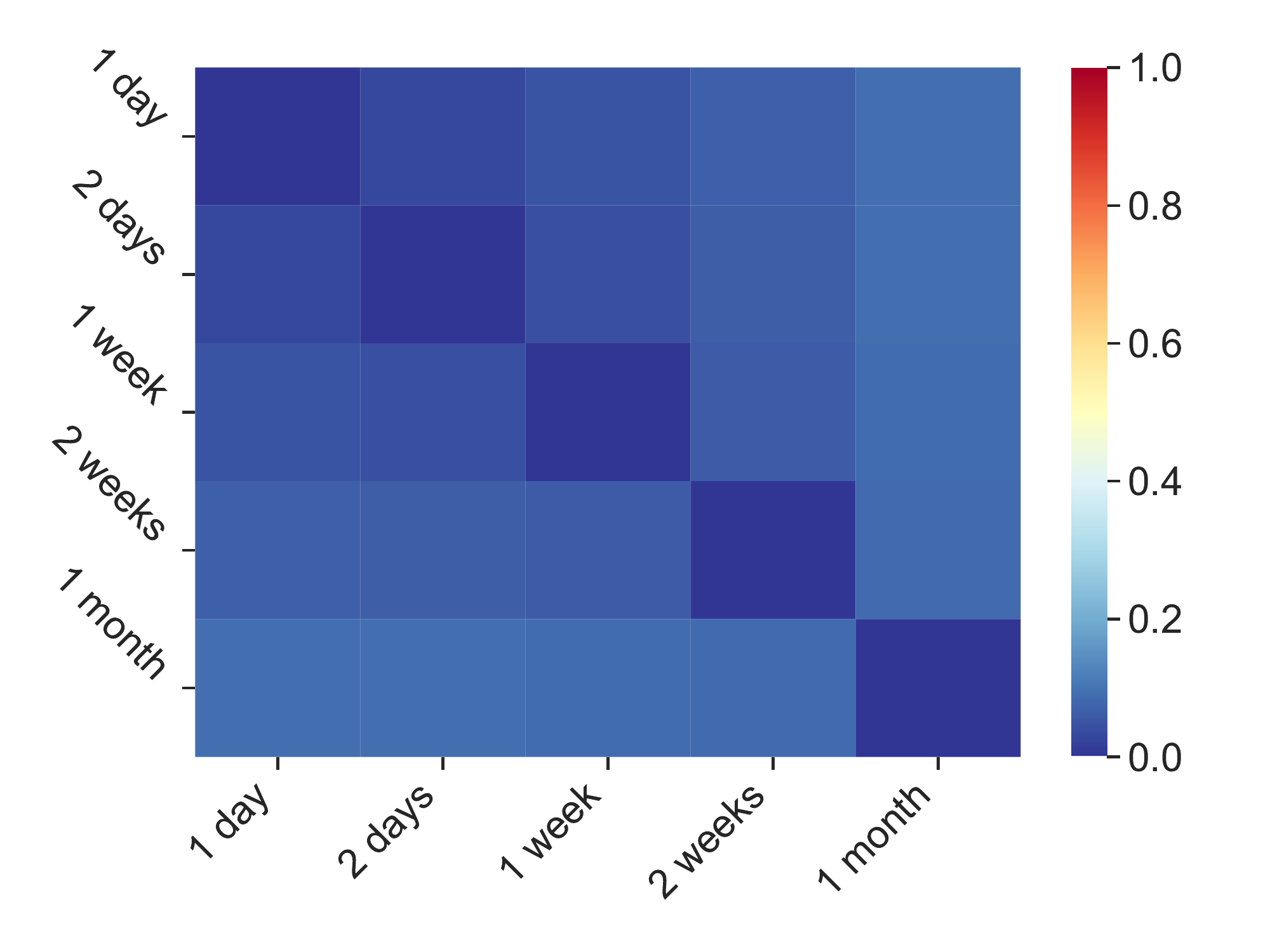}}
\end{minipage}
\caption{Left panel (\ref{resolution:a}): Divergence from original community structure for the five data sets of different sampling frequencies. It can be seen that the $VI$ is increasing steadily when moving from the finest to the coarsest resolution but does not exceed 10\%, indicating a high level of similarity between partitions. Right panel (\ref{resolution:b}): Heat map illustrating the value of $VI$ between each pair of the five data sets of different sampling frequencies. From the heat map, it is apparent that there is a considerable degree of consistency across sampling frequencies, but the similarity degrades steadily when moving from the finest to the coarsest resolution.}
\label{fig:resolution}
\end{figure}

\paragraph{Community detection: a temporal multiresolution analysis} Once the mesoscopic organisation of the CDS market has been detected, one may wonder how stable (i.e. `robust over time') this organisation is. This amounts at investigating the presence of discrepancies from the original community structure once the log-returns constituting the time series are sampled at different frequencies, e.g. weekly or monthly. For consistency with the results presented so far, the same period of ten years is considered but our analysis is now focused on the set of time series induced by the following array of time resolutions

\begin{equation}
\Delta_t \in \left\lbrace \mbox{1 day}, \mbox{2 days}, \mbox{1 week}, \mbox{2 weeks}, \mbox{1 month}\right\rbrace.
\end{equation}
 
To measure the effects of the temporal resolution, we employ the index known as \emph{variation of information} ($VI$), an information-theoretic measure quantifying the difference between any two partitions. Two different partitions can be represented by two $N$-dimensional vectors $\vec{\bm{\sigma}_1}$ and $\vec{\bm{\sigma}_2}$ whose $i$-th component $\sigma_i$ denotes the module to which node $i$ belongs. The (normalised) variation of information is defined as

\begin{equation}
VI(\vec{\bm{\sigma}_1}:\vec{\bm{\sigma}_2})=1-\displaystyle\frac{I(\vec{\bm{\sigma}_1}:\vec{\bm{\sigma}_2})}{H(\vec{\bm{\sigma}_1}:\vec{\bm{\sigma}_2})}
\end{equation}
where $I(\vec{\bm{\sigma}_1}:\vec{\bm{\sigma}_2})$ is the \emph{mutual information}, which is defined as follows

\begin{equation}
I(\vec{\bm{\sigma}_1}:\vec{\bm{\sigma}_2}) = \displaystyle\sum_{i=1}^{N}\displaystyle\sum_{j=1}^{N}p(\sigma_i^1,\sigma_j^2)\log\left[\displaystyle\frac{p(\sigma_i^1,\sigma_j^2)}{p(\sigma_i^1)p(\sigma_j^2)}\right]
\end{equation}
(with $p(\sigma_i^1)$ being the probability for a node to belong to $\sigma_i$ in partition 1, $p(\sigma_j^2)$ being the probability for a node to belong to $\sigma_j$ in partition 2 and $p(\sigma_i^1,\sigma_j^2)$ being the joint probability distribution for a node to belong to $\sigma_i$ in partition 1 \emph{and} to $\sigma_j$ in partition 2) and $H(\vec{\bm{\sigma}_1}:\vec{\bm{\sigma}_2})$ is the \emph{joint entropy}, which is defined as follows

\begin{equation}
H(\vec{\bm{\sigma}_1}:\vec{\bm{\sigma}_2})=\displaystyle\sum_{i=1}^{N}\displaystyle\sum_{j=1}^{N}p(\sigma_i^1,\sigma_j^2)\log\left[p(\sigma_i^1,\sigma_j^2)\right].
\end{equation}

Notice that, unlike mutual information, the variation of information is a true metrics since it satisfies the triangle inequality. The divergence from the community structure presented in \Cref{sec:results}, for each additional set of time series is depicted by \Cref{resolution:a}. From the chart, it is apparent that $VI$ is increasing steadily when moving from the finest to the coarsest resolution but does not exceed 10\% in any case, indicating a high level of similarity between partitions. The value of the $VI$ between each pair of data-sets is, instead, shown in \Cref{resolution:b}. This result provides some support for the conceptual premise that community structure does not (strongly) depend on the level of temporal resolution at which our data-set is considered.

\begin{figure}
\centering
\includegraphics[width=0.5\textwidth]{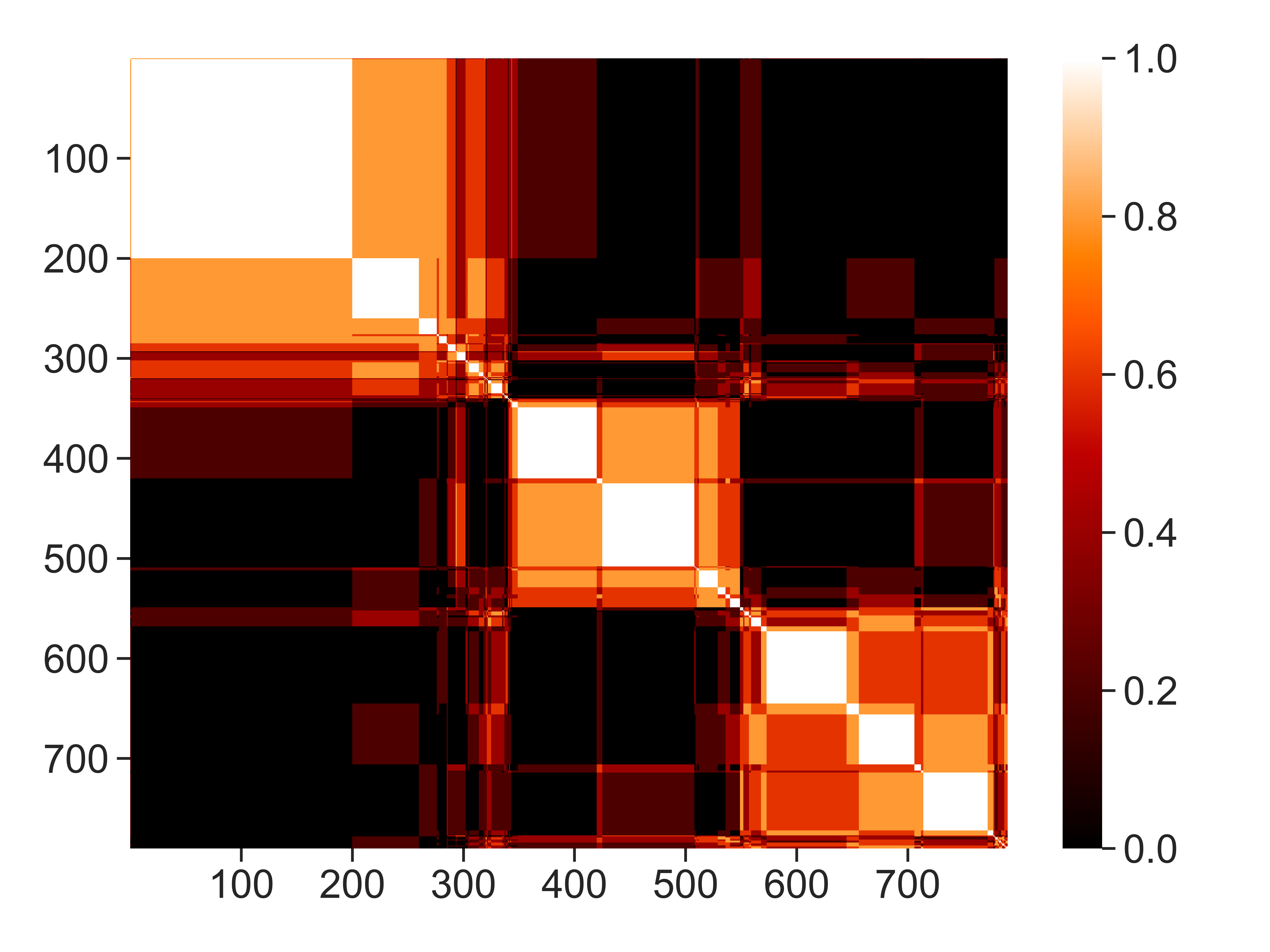}
\caption{Multifrequency heat map showing the normalised co-occurrence of different pairs of issuers within the same community, for the same time period but over various temporal resolutions of the original time series. The issuers have been ordered using hierarchical clustering with average linkage to position issuers with a high degree of co-occurrence next to each other.}
\label{fig:coherence_freq}
\end{figure}

The $VI$-based analysis contributes to our understanding of the robustness of the community structure with respect to different temporal resolutions; however, we would like to extend this kind of analysis at the issuer level, by assessing how robust is the assignment of issuers to communities by measuring the frequency with which any two issuers are assigned to the same community over different temporal partitions. The results of this analysis are presented in \Cref{fig:coherence_freq}. The heat map shows how frequently issuers co-occur within the same communities across the time resolutions considered here. In case two issuers are found within the same community for \emph{each} time resolution, the entry corresponding to the considered pair of issuers in the heat map is drawn in white, while if they are \emph{never} found within the same community the entry is drawn in black. As the heat map shows, three `hard cores' of issuers appear, indicating that the issuers belonging to them are assigned to the same community for the vast majority of the temporal resolutions; in addition, there are also few `soft issuers' who move across communities for different time resolutions, offering an explanation for the small variation observed in \Cref{fig:resolution}.

\paragraph{Community detection: temporal stability} One of the major challenges that managers measuring portfolio risk face is that of determining the appropriate period of history to use when estimating correlations to be used in their models. According to \cite{loretan2000iii}, using a relatively short period of data might be dangerous: in case the employed period is uncharacteristically stable, in fact, the estimated correlations may be lower than average, leading to excessive risk taking; on the other hand, if the used interval is relatively volatile, the resulting correlations can be unrealistically high, leading to excessive risk aversion. However, choosing a longer time series is not guaranteed to produce more reliable estimates: the ever-evolving nature of financial markets deems undesirable to rely on data from the distant past. In order to be confident that the results presented in \Cref{sec:results} can provide useful insights for risk managers, it is required to study the time dynamics of the detected communities.

% \begin{figure}
% \begin{center}
% \begin{minipage}{0.95\linewidth}
% \vspace*{0.5cm}
% \subfloat[]{
% \resizebox*{.49\linewidth}{!}{\includegraphics{graphics/IV_time.pdf}}\label{time:a}}
% \subfloat[]{
% \resizebox*{.49\linewidth}{!}{\includegraphics{graphics/IV_multires.pdf}}\label{time:b}}
% \caption{Left panel (\ref{time:a}): Divergence from initial community structure for the ensuing six-month windows. The values of $VI$ do not exceed 10\% for any of the six-month windows, while no particular trend can be observed. Right panel (\ref{time:b}): The heat map illustrates the value of $VI$ between every pair of six-month time windows, as well as the $VI$ between each window and the total ten-year period (rightmost column and bottom row). It can be seen that there is a high degree of community coherence over time.
% \label{fig:robustness_time}}
% \end{minipage}
% \end{center}
% \end{figure}

\begin{figure}
\centering
\begin{minipage}{.49\linewidth}
\centering
\vspace*{0.6cm}
\subfloat[]{\label{time:a}\includegraphics[width=\textwidth]{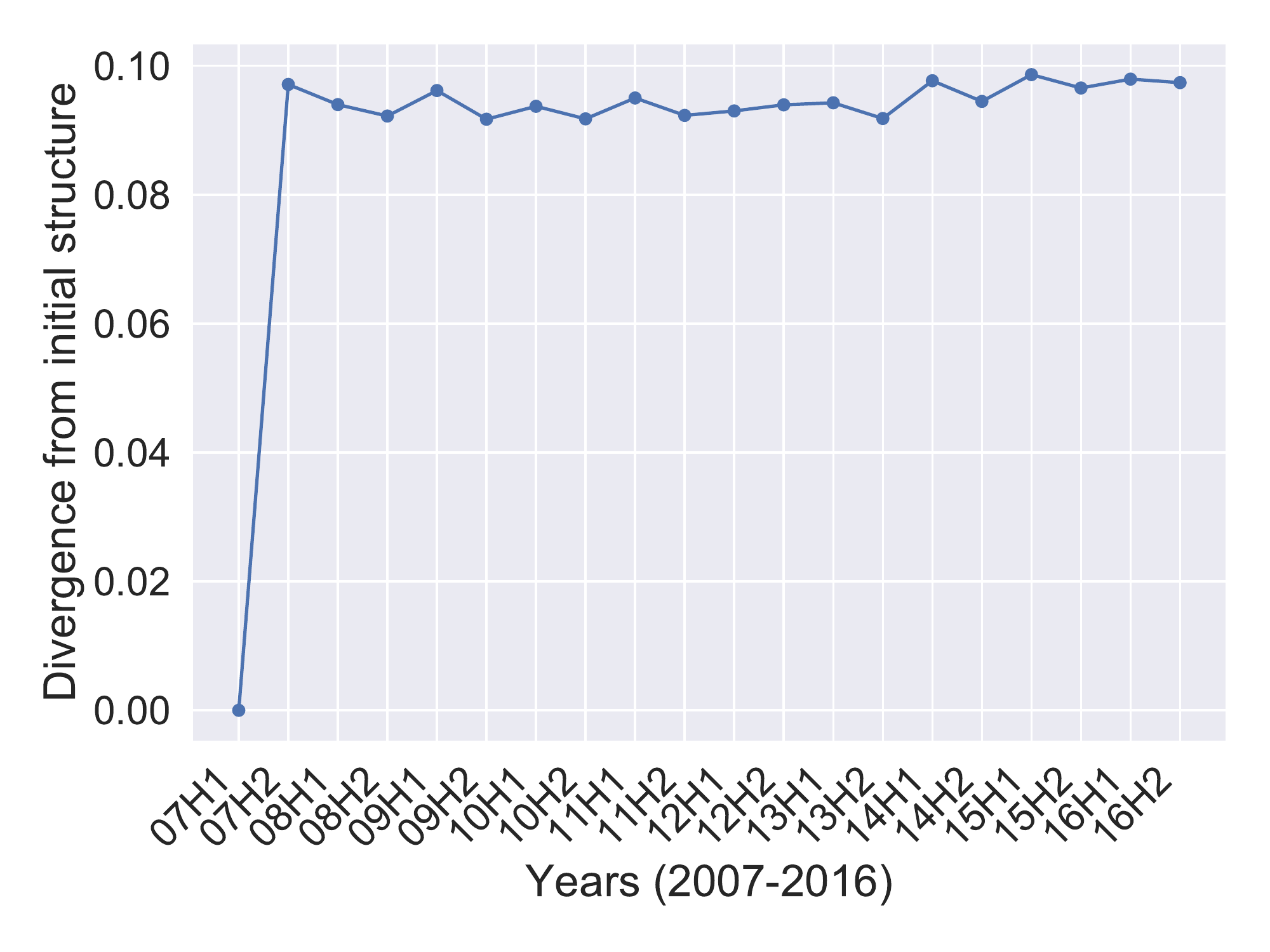}}
\end{minipage}
\begin{minipage}{.50\linewidth}
\centering
\subfloat[]{\label{time:b}\includegraphics[width=\textwidth]{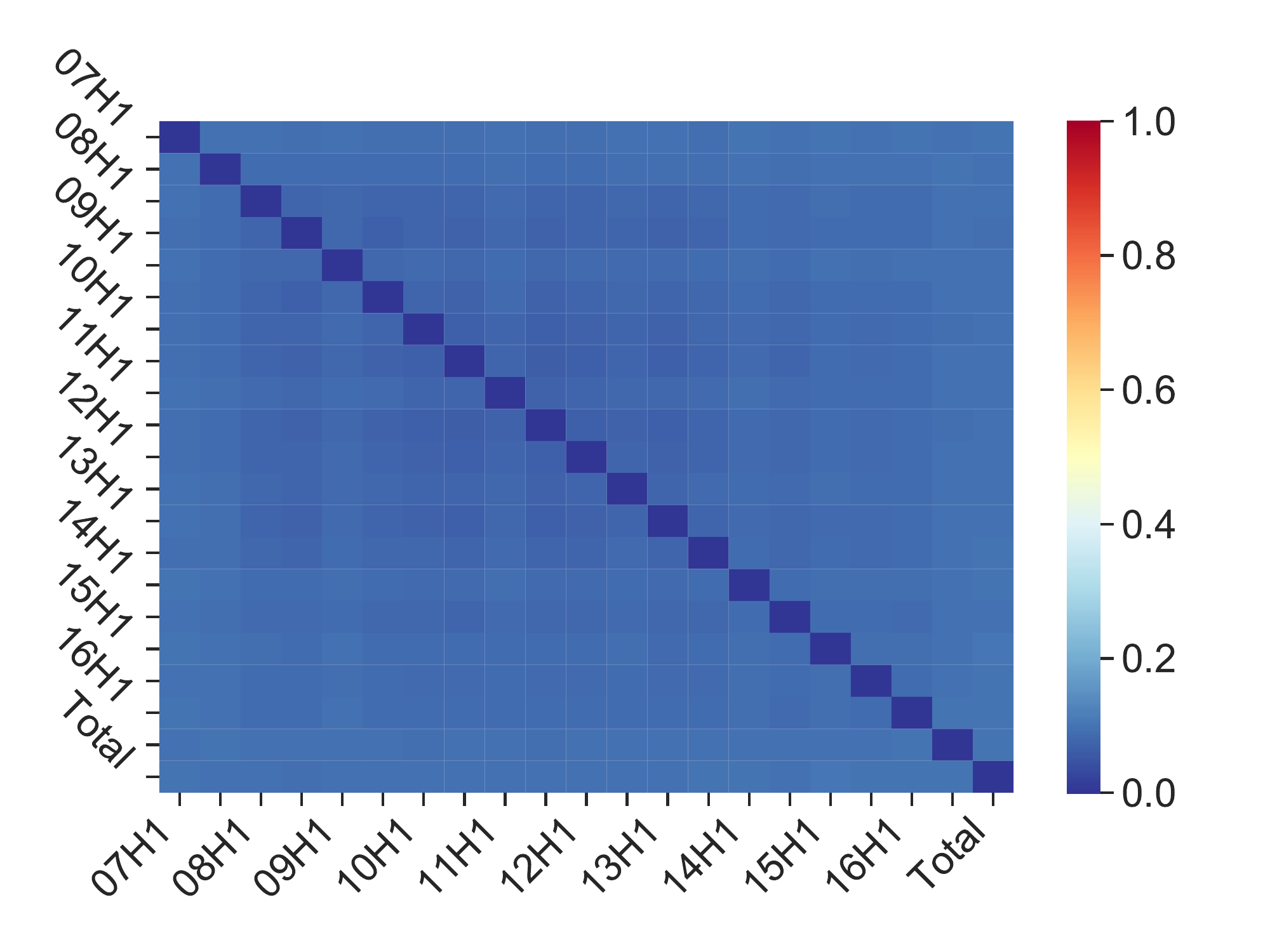}}
\end{minipage}
\caption{Left panel (\ref{time:a}): Divergence from initial community structure for the ensuing six-month windows. The values of $VI$ do not exceed 10\% for any of the six-month windows, while no particular trend can be observed. Right panel (\ref{time:b}): The heat map illustrates the value of $VI$ between every pair of six-month time windows, as well as the $VI$ between each window and the total ten-year period (rightmost column and bottom row). It can be seen that there is a high degree of community coherence over time.}
\label{fig:robustness_time}
\end{figure}

To analyse the stability of communities over the course of time, we employ a non-overlapping sliding window of six months. \Cref{time:a} illustrates the divergence from the community structure detected using the first six-month window throughout the years. It can be seen that there are no significant fluctuations, with $VI$ not exceeding 10\% for any of the six-month periods. Financial markets have the tendency to be more globally correlated during crisis periods, overriding the effect of group level correlations. Thus, a question that naturally arises is whether the evolution of the community structure is influenced by the financial crisis period of 2007-2009. We recall that the choice of $||C||$ in \Cref{eq:new_modularity} is already discounting the evolution of the volatility of the market. However, it is still interesting to see that the values of $VI$ remain low and stationary. This indicates that, regardless of the market stress between 2007-2009, the composition of the communities has remained stable over time. To further improve our understanding on the community coherence, in \Cref{time:b} we provide a heat map showing the mutual $VI$ between every two pairs of six-month windows. Each square in the matrix represents the value of $VI$ between the $i$-th and $j$-th six-month period, while the last row and column represent the value of $VI$ between each six-month period and the partition obtained using the full ten-year sample. The results indicate that there was little movement of issuers between communities during the ten-year period. In addition, it can be seen that there is little difference between these periods and the community structure obtained when we using the entire ten-year period.

\begin{figure}
\centering
\includegraphics[width=0.5\textwidth]{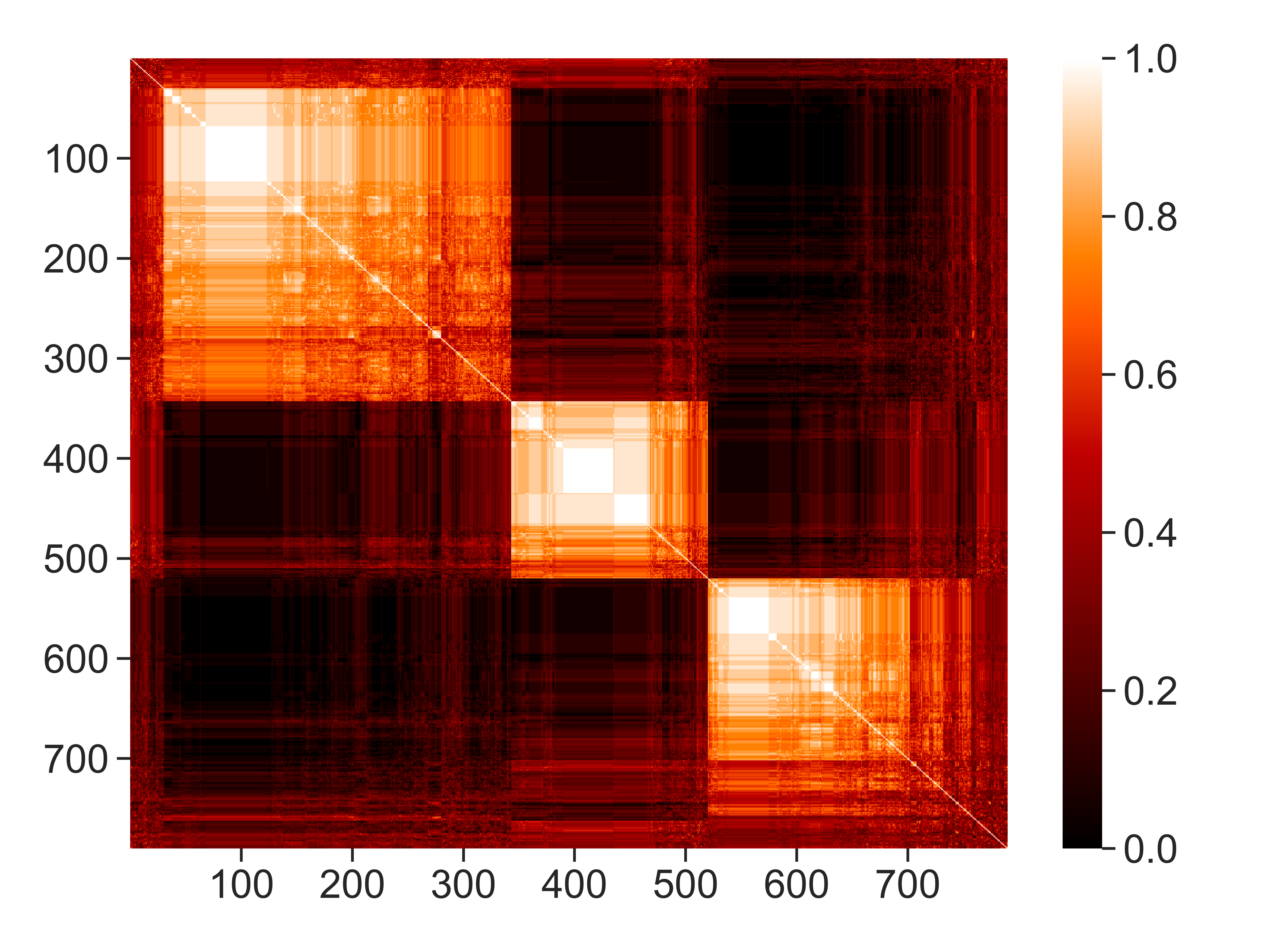}
\caption{Coherence of communities over time. The heat map shows the frequency of co-occurrence of different pairs of issuers within the same community over time. The communities appear to be tight-knit and unwavering, maintaining coherence over the course of the ten-year time frame. The issuers have been ordered using hierarchical clustering with average linkage to position issuers with a high degree of co-occurrence next to each other.}
\label{fig:coherence_time}
\end{figure}

Finally, having determined that the communities do not exhibit significant fluctuations over time, we use the same sliding window to examine the stability of communities over time at the issuer level. In a similar fashion to the temporal multiresolution analysis, we plot a heat map containing the frequency with which each pair of issuers can be found within the same community over the course of the ten-year time frame. As \Cref{fig:coherence_time} shows, pairs of issuers appearing to be in the same community for all the six-month windows have white entries in the heat-map, while the entries corresponding to pairs of issuers never appearing within the same community are drawn in black. After a closer inspection, it can be seen that the three communities detected by using the full ten-year history appear to be tight-knit and unwavering, thus maintaining a high degree of coherence over the course of time. A small number of issuers moving fluidly from one community to another can still be appreciated.

\section{Default risk charge model}\label{sec:model}

\subsection{Model specification}\label{subsec:model_spec}

Consider a portfolio of $m$ issuers, indexed by $i=1, 2\dots m$ and a fixed time horizon of $T=1$ year. The overall portfolio loss is modelled by a random variable $L$, defined as the sum of the individual losses on issuers' default, i.e. $L=\sum_{i=1}^{m}L_i$, with

\begin{equation}
L_i=q_ie_iY_i
\end{equation}
where $L_i = q_i e_i Y_i$ denotes the loss on issuer $i$, with $e_i$ and $q_i$ being, respectively, the \textit{exposure at default} and the \textit{loss given default} of issuer $i$ and $Y_i$ being the random default indicator taking the value 1 if issuer $i$ defaults before time $T$ and 0 otherwise. In order to define the probability distributions of $L_i$'s, as well as their dependence structure, we rely on a factor model approach, descending from the structural model of Merton \citep{merton1974pricing}, which is widely used for portfolio default risk modelling by regulators and financial institutions alike. Notable examples of this approach include the Asymptotic Single Risk Factor (ASRF) model \citep{gordy2003risk}, which is at the heart of Basel II credit risk capital charge, as well as industrial adaptations of Merton model such as the CreditMetrics \citep{morgan1997creditmetrics} and KMV models \citep{Kealhofer2001, Crosbie2002}.

Let us introduce a random variable $X_i$ representing issuer $i$'s creditworthiness. In the same spirit of Merton’s structural model, we specify that default occurs before time $T$ if the value of $X_i$ lies below a threshold $d_i$, or equivalently:

\begin{equation}
Y_i := \mathbf{1}_{[-\infty,d_i]}\left(X_i\right),
\end{equation}
where $\mathbf{1}_A(\cdot)$ is the indicator function of set $A$. Hence, modelling $Y_i$'s boils down to model the creditworthiness indices $X_i$ with $i=1\dots m$, which are linearly dependent on a vector $\mathbf{F}$ of $p<m$ systematic factors satisfying $\mathbf{F}\sim N_p(0,\Omega)$. Issuer $i$'s creditworthiness index is assumed to be driven by an issuer-specific combination $\tilde{F}_i = \boldsymbol{\alpha}_i^\intercal\mathbf{F}$ of the systematic factors:

\begin{eqnarray}
X_i = \sqrt{\beta_i}\tilde{F}_i + \sqrt{1-\beta_i}\epsilon_i{,}\label{eq:kernel}
\end{eqnarray}
where $\tilde{F}_i$ and $\epsilon_1\dots\epsilon_m$ are independent, standard normal variables (i.e. $\tilde{F}_i \sim N(0,1),\:\forall\:i$ and $\epsilon_i\sim N(0,1),\:\forall\:i$) and the latter ones model the idiosyncratic risk. The coefficient $\beta_i$ can be seen as a measure of sensitivity of $X_i$ to systematic risk, as it represents the proportion of the $X_i$ variation that is explained by the systematic factors. {The assumption that $\mbox{Var}[\tilde{F}_i]=1$ implies that $\boldsymbol{\alpha}_i^\intercal\Omega \boldsymbol{\alpha}_i=1$ for all $i$.} The correlations between asset returns are given by
\begin{equation}
\begin{split}
\rho(X_i,X_j)&=\mbox{Cov}[X_i,X_j]\\&= \left(1 - \beta_i\right)\mathbf{1}_{\lbrace i=j\rbrace} +\sqrt{\beta_i \beta_j}\mbox{Cov}[\tilde{F}_i, \tilde{F_j}]\\
&= \left(1 - \beta_i\right)\mathbf{1}_{\lbrace i=j\rbrace} + \sqrt{\beta_i \beta_j}\boldsymbol{\alpha}_i^\intercal\Omega \boldsymbol{\alpha}_j
\end{split}
\end{equation}
(since $\tilde{F}_i$ and $\epsilon_1\dots\epsilon_m$ are independent, standard normal variables and $\mbox{Var}[X_i]=1$). In order to set up the model we need to determine $\boldsymbol{\alpha}_i$ and $\beta_i$ for each issuer and $\Omega$ (while ensuring that $\boldsymbol{\alpha}_i^\intercal\Omega\boldsymbol{\alpha}_i=1$).

We choose $d_i$ such that $\mathbb{P}(Y_i =1)=p_i$, where $p_i$ is the marginal \textit{probability of default} of issuer $i$. As a result, $d_i = F_{X_i}^{-1}(p_i)$, with $F_{X_i}(\cdot)$ being the cumulative distribution function of $X_i$. Given the normality of $X_i$, it follows that $d_i = \Phi^{-1}(p_i)$, with $\Phi^{-1}(\cdot)$ denoting the standard normal cumulative distribution function. The portfolio loss can, then, be written as follows:

\begin{equation}
L=\sum_{i=1}^{m} q_i e_i \mathbf{1}_{[-\infty,\Phi^{-1}(p_i)]}\left(\sqrt{\beta_i}\tilde{F}_i+\sqrt{1-\beta_i}\epsilon_i\right).
\end{equation}

For $p=1$, the specification above is equivalent to the ASRF model. In this model, the single systematic factor affecting all issuers is usually interpreted as the state of the economy and the correlation coefficients are regulatory prescribed. In multi-factor models ($p\geq 2$), latent or observable factors encompassing regional or industry characteristics are typically used by modellers to capture the portfolios correlation structure.

\subsection{Model calibration}\label{subsec:calibration}
Models used by banks for DRC calculations are required to account for systematic risk via multiple systematic factors of two different types \cite[Paragraph~186(b)]{basel2016minimum}. For the first systematic factor, we consider a global factor that is common to all issuers, reflecting the overall state of the economy. We adopt this approach due to relevant literature suggesting strong dependence of changes in default risk on global effects \citep{aretz2013common}. For the second systematic factor we consider factors representing industry and region effects, as well as community and subcommunity effects. Even though a model with three types of systematic factors would not be in line with the regulatory requirements for the calculation of DRC, for comparison purposes we also consider a model with global, industry, and region systematic factors, an approach commonly adopted in the industry for the calculation of IRC.

In addition to the types of systematic factors, the regulatory rule-set specifies that correlations must be calibrated using credit spreads or listed equity prices over a period of at least ten years that includes a period of stress. We calibrate the model presented in \Cref{subsec:model_spec} using CDS spreads covering the period between 1 January 2007-31 December 2016 which includes the `stressed' period between 2007 and 2009. In our model setting, the liquidity horizon is set as one year and, as a result, the correlations should be
measured over the same horizon. However, non-overlapping annual log-returns from ten years history contain only nine points, which is not sufficient to yield a reliable correlation estimate. If overlapping log-returns are considered alternatively, the sample size can be sufficiently big, but the linear relation between the data series can be distorted. Moreover, certain bias can be introduced into the correlation estimate via auto-correlation, which usually leads to over-estimations. Instead of using overlapping annual returns, we chose to use to monthly non-overlapping returns over the ten-year period. This approach leads to a sufficient number of data points for correlation estimation and can be seen as a reasonable compromise. The implied hypothesis is that correlations measured over monthly and annual horizons are interchangeable and can be used as a predictor for future one-year correlations. According to \cite{wilkens2017default}, this assumption can be questioned, but, it is hard to reject from a statistical point of view, if one takes into account the uncertainty of the correlation measurement itself.

 We start by scaling each individual time series to have zero mean and unit variance. At each time point, global $(X_{G,t})$, industry $(X_{I(j),t})$, region $(X_{R(k),t})$, community $(X_{C(l),t})$ and subcommunity $(X_{S(n),t})$ returns are derived from the corresponding cross-section of the issuer returns. All the resulting factor time series have zero mean. The dependence of the region, industry, community and subcommunity factors on global returns is explored by running the following linear regression models

\begin{equation}
\label{eq:regression1}
\begin{split}
&X_{I(j),t}=\gamma_{I(j)}X_{G,t}+\varepsilon_{I(j),t},\\
&X_{R(k),t}=\gamma_{R(k)}X_{G,t}+\varepsilon_{R(k),t},\\
&X_{C(l),t}=\gamma_{C(l)}X_{G,t}+\varepsilon_{C(l),t},\\
&X_{S(n),t}=\gamma_{S(n)}X_{G,t}+\varepsilon_{S(n),t}
\end{split}
\end{equation}
where $\gamma_{I(j)}$, $\gamma_{R(k)}$, $\gamma_{C(l)}$ and $\gamma_{S(n)}$ are coefficients weighing the global factor and $\varepsilon_{I(j),t}$, $\varepsilon_{R(k),t}$, $\varepsilon_{C(l),t}$ and $\varepsilon_{S(n),t}$ are the industry-, region-, community- and subcommunity-specific residuals, respectively.  The full regression results on the basis of \Cref{eq:regression1} for the period between January 2007 to December 2016 can be found in \Cref{tab:results} in \Cref{app:reg_results}. The majority of the factor returns move in line with the global returns, with coefficients not significantly different from one. In addition, the proportion of variance explained by the global returns is high (as indicated by the values of $R^2$, between 63\% and 96\%), highlighting the leading role of the global factor.

\begin{table}
\begin{center}
\begin{minipage}{\textwidth}
\tbl{The table provides the statistics for the $R^2$ between the individual issuers and systematic factor(s). The model estimation is based on the following settings: global factor only (Model 1); global and industry factors (Model 2); global and region factors (Model 3); global, region, and industry factors (Model 4); global and community factors (Model 5); and global and subcommunity factors (Model 6). Industry, region, community, and subcommunity factors are defined as cross-sectional averages at each time point and taken as already decomposed into a global factor and residuals. The estimation is based on non-overlapping monthly log-returns and conducted from January 2007 to December 2016, which was identified as a recent 10-year period which included the `stressed' period between 2007 and 2009.}
{\begin{tabular}{@{}ccccccc}\toprule
Individual $R^2$ & Model 1:   & Model 2: & Model 3: & Model 4: & Model 5: & Model 6:\\                             versus                &  Global &  Global and  &  Global and &  Global, region, &  Global and &  Global and \\
 systematic & factor & industry & region & and industry & community & subcommunity\\
 factors & only & factors & factors & factors & factors & factors\\
\colrule
Average	&	48.1\%	&	54.5\%	&	54.2\%	&	58.1\%	&	55.4\%	&	58.2\%	\\
SD	&	15.8\%	&	16.4\%	&	17.6\%	&	17.5\%	&	16.7\%	&	18.0\%	\\
Minimum	&	0.0\%	&	0.6\%	&	0.0\%	&	1.4\%	&	1.4\%	&	0.8\%	\\
Maximum	&	0.82\%	&	84.1\%	&	96.3\%	&	96.5\%	&	88.4\%	&	92.6\%	\\
\botrule
\end{tabular}}
\label{tab:rsquared}
\end{minipage}
\end{center}
\end{table}

% \begin{table}[t!]
% \centering
% \caption{The table provides the statistics for the $R^2$ between the individual issuers and systematic factor(s). The model estimation is based on the following settings: global factor only (Model 1); global and industry factors (Model 2); global and region factors (Model 3); global, region, and industry factors (Model 4); global and community factors (Model 5); and global and subcommunity factors (Model 6). Industry, region, community, and subcommunity factors are defined as cross-sectional averages at each time point and taken as already decomposed into a global factor and residuals. The estimation is based on non-overlapping monthly log-returns and conducted from January 2007 to December 2016, which was identified as a recent 10-year period which included the `stressed' period between 2007 and 2009.}\label{tab:rsquared}
% \begin{tabular}{ccccccc}

% \hline
% \\
% Individual $R^2$ & Model 1:   & Model 2: & Model 3: & Model 4: & Model 5: & Model 6:\\                             versus                &  Global &  Global and  &  Global and &  Global, region, &  Global and &  Global and \\
%  systematic & factor & industry & region & and industry & community & subcommunity\\
%  factors & only & factors & factors & factors & factors & factors\\
%  \\
% \hline
% \\
% Average	&	48.1\%	&	54.5\%	&	54.2\%	&	58.1\%	&	55.4\%	&	58.2\%	\\
% SD	&	15.8\%	&	16.4\%	&	17.6\%	&	17.5\%	&	16.7\%	&	18.0\%	\\
% Minimum	&	0.0\%	&	0.6\%	&	0.0\%	&	1.4\%	&	1.4\%	&	0.8\%	\\
% Maximum	&	0.82\%	&	84.1\%	&	96.3\%	&	96.5\%	&	88.4\%	&	92.6\%	\\
% \\ \hline

% \end{tabular}
% \end{table}

Turning now to the case of a single issuer, it is important to note that by regressing the industry $(X_{I(j)})$, region $(X_{R(k)})$, community $(X_{C(l)})$ and subcommunity $(X_{S(n)})$ returns against the global returns $(X_{G})$, we have essentially orthogonalised the rest of the factors relative to the global factor. Hence,  in addition to the global returns $(X_{G})$, we use the residuals $\varepsilon_{I(j)}, \varepsilon_{R(k)},\varepsilon_{C(l)}$, and  $\varepsilon_{S(n)} $ from \cref{eq:regression1} as explanatory factors for the returns of a single issuer, representing industry, region, community, and subcommunity effects respectively. In the following we discuss the calibration of a model with a global and a subcommunity factor; calibration of other model variants should then be straightforward.  Recall that issuer $i$'s creditworthiness index $X_i$ follows the dynamics presented in \Cref{eq:kernel}. In our case, $\tilde{F_i}:=\alpha_{G(i)}X_G+\alpha_{S(i)}\epsilon_{S(i)}$, where the coefficients $\alpha_{G(i)}$ and $\alpha_{S(i)}$ have been rescaled so that $\tilde{F_i}\sim N(0,1)$, i.e.

\begin{equation}
\alpha_{G(i)}:= \frac{\hat\alpha_{G(i)}}{\Psi_i},\qquad\alpha_{S(i)}:= \frac{\hat\alpha_{S(i)}}{\Psi_i}
\end{equation}
with $\hat\alpha_{G,i}$ and $\hat\alpha_{S,i}$ being the factor loadings of the regression model
\begin{equation}
X_{i,t} = \hat\alpha_{G(i)}X_{G,t}+  \hat\alpha_{S(i)}\epsilon_{S(i),t} + \varepsilon_{i,t}
\label{eq:regression2}
\end{equation}
and $\Psi_i=\sigma[\hat\alpha_{G(i)}X_G+\hat\alpha_{S(i)}\epsilon_{S(i)}]$. Thus, we calibrate the factor loadings $\hat\alpha_{G(i)}$ and $\hat\alpha_{S(i)}$ by running the above regression. If we collect $X_G$ and $\epsilon_{S(i)}$ into a matrix $F_i$, then the least squares estimator $\hat{\boldsymbol{\alpha}}_i$ equalises the two sides of the following equation:

\begin{equation}
X_i=F_i\hat{\boldsymbol{\alpha}}_i
\end{equation}
or, in other words, $\hat{\boldsymbol{\alpha}}_i,\:\forall\:i$ represent estimates of the coefficients appearing in \Cref{eq:regression2}:

\begin{equation}
\hat{\boldsymbol{\alpha}}_i=(F^\intercal_iF_i)^{-1}(F^\intercal_iX_i).
\end{equation}
Finally, the coefficient $\beta_i$ from \Cref{eq:kernel} is the $R^2$ of this regression, representing the proportion of variance for $X_i$ explained by the systematic factors.

We estimate the parameters of six model variants on the basis of the calibration process described previously: global factor only (Model 1); global and industry factors (Model 2); global and region factors (Model 3); global, region, and industry factors (Model 4); global and community factors (Model 5) and global and subcommunity factors (Model 6). The statistics for the individual $R^2$ versus systematic factors are compared in \Cref{tab:rsquared}. As it can be seen from the table, not surprisingly, the model based only on the global factor provides the worst fit to the data with an average $R^2$ of 48.2\%. After the introduction of the industry factor the average $R^2$ increases to 54.5\%. A comparable value (54.2\%) is obtained if instead of the industry factor we introduce a region factor. The model based on global and community factors (Model 5) provides a slightly better fit than the other two-factor models with an average $R^2$ of 55.4\%. The best fit is achieved by the model based on global and subcommunity factors (Model 6) with an average $R^2$ of 58.2\%, outperforming even the three-factor model based on global, region, and industry factors. This result is particularly interesting since the difference in the average $R^2$ between Model 6 and the other two-factor models is comparable to the difference between the other two-factor models and the model based only on the global factor.

\begin{figure}
\centering
\includegraphics[width=0.8\textwidth]{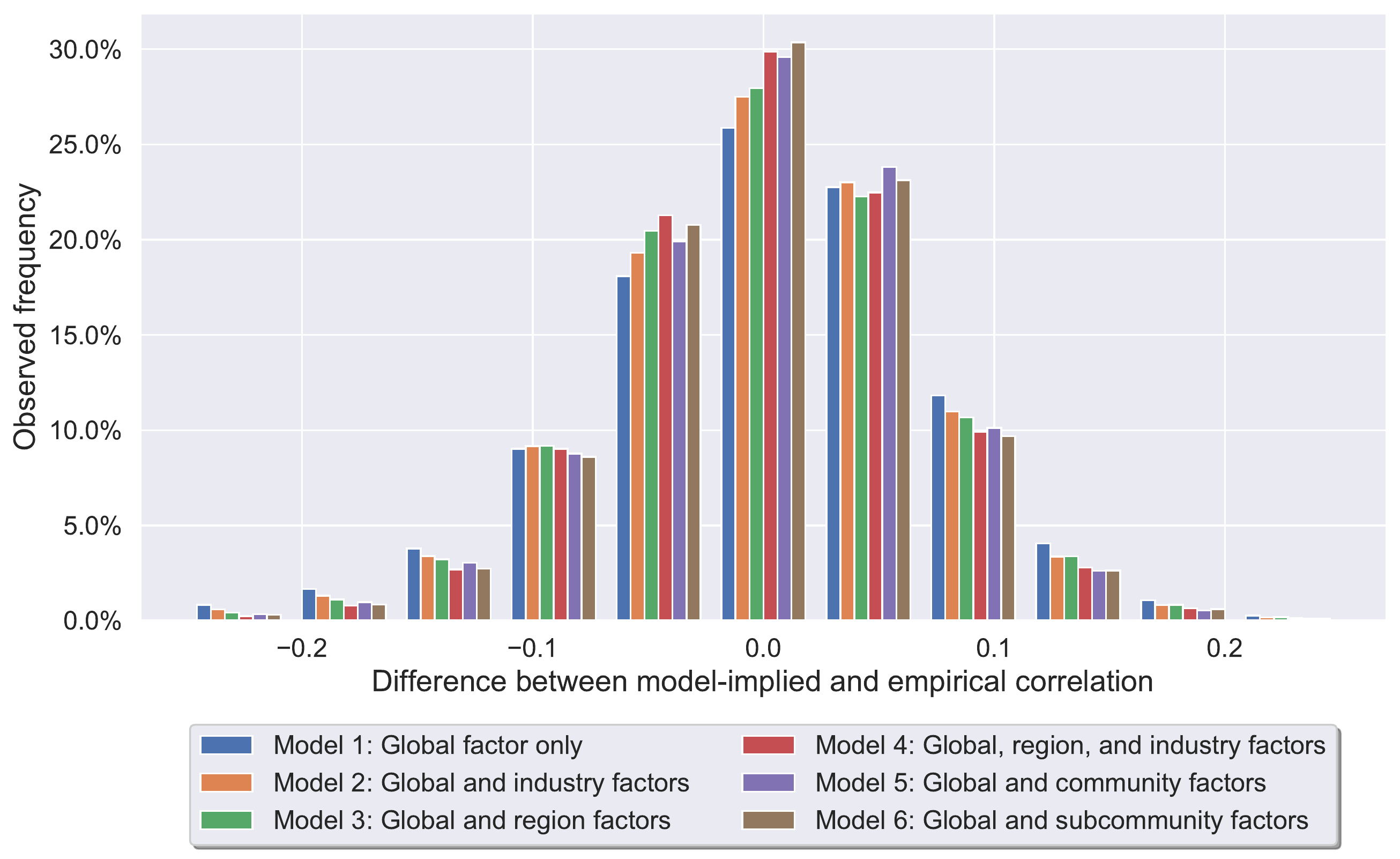}
\caption{The figure shows the distribution of the differences between the model-implied and empirical (pairwise) correlations. As for the model, the six variants (global factor only, global/industry, global/region, global/industry/region, global/community, global/subcommunity factors) are
explored.}
\label{fig:cor_diff}
\end{figure}

Having estimated the factor loadings and covariance matrices for the six model variants, we are able to obtain the distribution of the differences between the model-implied and empirical (pairwise) correlations (shown in \Cref{fig:cor_diff}): after close inspection, it becomes evident that the distributions of the correlation errors based on Model 4 and Model 6 are heavier around zero and have thinner tails. The distribution of the actual empirical correlations is shown in \Cref{fig:emp_cor} in \Cref{sec:emp_correlations}.

\subsection{Numerical experiments}

In order to study the properties of the framework presented in \Cref{sec:model} we set up four synthetic test portfolios:

\begin{itemize}
\item \textbf{Portfolio A}: Long-only portfolio consisting of 36 sovereign issuers from the iTraxx SovX index family.
\item \textbf{Portfolio B}: Long-only portfolio consisting of 89 corporate issuers (Financials and Non-Financials) from the iTraxx Europe index.
\item \textbf{Portfolio C}: Long-only consisting of 125 issuers from Portfolio A and Portfolio B combined.
\item \textbf{Portfolio D}: Long-short portfolio consisting of 22 long positions on issuers from Financials and 22 short positions on issuers from Non-Financials from the iTraxx Europe index selected such that the average default probability between the two groups is the same.
\end{itemize}

For the purposes of our numerical experiments, we use historical default rates per rating from \cite{sp2019a} and \cite{sp2019b} as \textit{probabilities of default}. The historical defaults rates per rating are shown in \Cref{tab:probabilities}. In accordance with regulatory requirements \cite[Paragraph 186(b)]{basel2016minimum}, probabilities of default are subject to a floor of 3 bps. The detailed composition of the synthetic portfolios in terms of rating, as well as the mean and standard deviation of the corresponding probabilities of default are shown in \Cref{tab:portfolio_ratings}.

\begin{table}
\begin{center}
\begin{minipage}{\textwidth}
\tbl{The table provides
the historical default rates (\%) per rating. Source \cite{sp2019a,sp2019b}}
{\begin{tabular}{@{}rrr}\toprule
\multirow{2}{*}{Rating}	&	\multicolumn{1}{c}{Corporates}	&		\multicolumn{1}{c}{Sovereigns}	\\
	&		\multicolumn{1}{c}{(1981-2018)}	&		\multicolumn{1}{c}{(1975-2018)}	\\
\colrule
AAA	&	0.00	&	0.00	\\
AA	&	0.02	&	0.00	\\
A	&	0.06	&	0.00	\\
BBB	&	0.17	&	0.00	\\
BB	&	0.65	&	0.49	\\
B	&	3.44	&	2.82	\\
CCC/C	&	26.63	&	41.56	\\
\botrule
\end{tabular}}
\label{tab:probabilities}
\end{minipage}
\end{center}
\end{table}

% \begin{table}[t!]
% \centering
% \caption{The table provides
% the historical default rates per rating. Source \cite{sp2019a,sp2019b}}\label{tab:probabilities}
% \begin{tabular}{crr}

% \hline
% \\
% \multirow{2}{*}{Rating}	&	\multicolumn{1}{c}{Corporates}	&		\multicolumn{1}{c}{Sovereigns}	\\
% 	&		\multicolumn{1}{c}{(1981-2018)}	&		\multicolumn{1}{c}{(1975-2018)}	\\ \\
% \hline \\
% AAA	&	0.00	&	0.00	\\
% AA	&	0.02	&	0.00	\\
% A	&	0.06	&	0.00	\\
% BBB	&	0.17	&	0.00	\\
% BB	&	0.65	&	0.49	\\
% B	&	3.44	&	2.82	\\
% CCC/C	&	26.63	&	41.56	\\
% \\ \hline
% \end{tabular}
% \end{table}

As far as the \textit{exposure at default} is concerned, for the long-only portfolios we consider a constant and equally weighted exposure for each issuer such that $e_i = 1/m$ for all $i=1\dots m$ and $\sum_{i=1}^{m}=1$. For the long-short portfolio we consider $e_{i\in Financials}=1/22$ and $e_{i\notin Financials}=-1/22$, and as a result $\sum_{i=1}^{m}e_i=0$. Finally, for the sake of simplicity, the \textit{loss given default} parameter is set to 1 for all issuers, i.e. $q_i=1,\:i=1\dots m$. 

\begin{table}
\begin{center}
\begin{minipage}{\textwidth}
\tbl{The table provides
the composition of the synthetic test portfolios in terms of rating, as well as the mean and standard deviation for the corresponding probabilities of default.}
{\begin{tabular}{@{}cccccccc}\toprule
&AAA& AA & A& BBB& BB& Average PD& SD PD\\
\colrule
Portfolio A&4&7&6&14&5&0.09\%&0.16\%\\
Portfolio B&-&6&32&51&-& 0.12\%& 0.05\%\\
Portfolio C&4&13&38&65&5& 0.11\%& 0.10\%\\
Portfolio D&-&6&30&8&-& 0.08\% &0.05\%\\
\botrule
\end{tabular}}
\label{tab:portfolio_ratings}
\end{minipage}
\end{center}
\end{table}

% \begin{table}[b!]
% \centering
% \caption{The table provides
% the composition of the synthetic test portfolios in terms of rating, as well as the mean and standard deviation for the corresponding probabilities of default.}\label{tab:portfolio_ratings}
% \begin{tabular}{cccccccc}
% \hline
% \\
% &AAA& AA & A& BBB& BB& Average PD& SD PD\\ \\
% \hline \\
% Portfolio A&4&7&6&14&5&0.09\%&0.16\%\\
% \\
% \hline \\
% Portfolio B&-&6&32&51&-& 0.12\%& 0.05\%\\
% \\
% \hline \\
% Portfolio C&4&13&38&65&5& 0.11\%& 0.10\%\\
% \\
% \hline \\
% Portfolio D&-&6&30&8&-& 0.08\% &0.05\%\\
% \\ \hline
% \end{tabular}
% \end{table}

We, then, generate portfolio loss distributions and derive the associated risk measures by means of Monte Carlo simulations. This process entails generating joint realizations of the systematic and idiosyncratic risk factors and comparing the resulting critical variables with the corresponding default thresholds. By this comparison, we obtain the default indicator $Y_i$ for each issuer and this enables us to calculate the overall portfolio loss for this trial. A liquidity horizon of 1 year is assumed throughout and the results are based on calibrations according to \Cref{subsec:calibration} and simulations with ten million sample paths each. For a given confidence level $\alpha\in [0,1]$, the $\mbox{VaR}_{\alpha}$ is defined as the $\alpha$-quantile of the loss distribution:

\begin{equation}
\mbox{VaR}_{\alpha} (L) = \inf\lbrace l\in \mathbb{R}:\mathbb{P}(L\leq l)\geq \alpha)\rbrace.
\end{equation}

\begin{table}
\begin{center}
\begin{minipage}{\textwidth}
\tbl{The table provides
the quantiles of the loss distribution for the corresponding portfolios and for each of the six model configurations.}
{\begin{tabular}{@{}cccccccc}\toprule
&\multirow{4}{*}{$\alpha$} & Model 1:   & Model 2: & Model 3: & Model 4: & Model 5: & Model 6:\\                                           &  &  Global &  Global and  &  Global and &  Global, region, &  Global and &  Global and \\
 & & factor & industry & region & and industry & community & subcommunity\\
&  & only & factors & factors & factors & factors & factors\\
\colrule
Portfolio A:&0.99& 2.8\%& 2.8\%& 2.8\%& 2.8\%& 2.8\%& 2.8\%\\
Sovereign bonds, &0.995& 2.8\% & 5.6\%& 5.6\%& 5.6\%& 2.8\%& 2.8\%\\
long position&0.999 &8.3\% &11.1\%& 11.1\%& 11.1\%& 8.3\%& 11.1\%\\
\colrule
Portfolio B:&0.99& 3.4\%& 3.4\%& 2.2\%& 2.2\%& 3.4\%& 3.4\%\\
Corporate bonds,&0.995& 5.6\%& 5.6\%& 5.6\%& 5.6\%& 5.6\%& 5.6\%\\
long position&0.999 &14.6\%& 14.6\%& 16.9\%& 16.9\%& 15.7\%& 16.9\%\\
\colrule
Portfolio C:&0.99& 2.4\%& 2.4\%& 2.4\%& 2.4\%& 2.4\%& 2.4\%\\
Combination of&0.995& 4.8\%& 4.8\%& 4.8\%& 4.8\%& 4.8\%& 4.8\%\\
portfolios A and B&0.999& 12.8\%& 12.8\%& 13.6\%& 13.6\%& 12.8\%& 13.6\%\\ 
\colrule
Portfolio D:&0.99& 0.0\%& 0.0\%& 0.0\%& 0.0\%& 0.0\% &0.0\%\\
Corporate bonds,&0.995& 4.5\%& 4.5\%& 4.5\%& 0.0\%& 4.5\%& 4.5\%\\
long/short position&0.999& 9.1\%& 13.6\% &9.1\% & 9.1\% & 13.6\% & 13.6\%\\
\botrule
\end{tabular}}
\label{tab:losses}
\end{minipage}
\end{center}
\end{table}

% \begin{table}[t!]
% \centering
% \caption{The table provides
% the quantiles of the loss distribution for the corresponding portfolios and for each of the six model configurations.}\label{tab:losses}
% \begin{tabular}{cccccccc}
% \hline
% \\
% &\multirow{4}{*}{$\alpha$} & Model 1:   & Model 2: & Model 3: & Model 4: & Model 5: & Model 6:\\                                           &  &  Global &  Global and  &  Global and &  Global, region, &  Global and &  Global and \\
%  & & factor & industry & region & and industry & community & subcommunity\\
% &  & only & factors & factors & factors & factors & factors\\
%  \\
% \hline \\
% Portfolio A:&0.99& 2.8\%& 2.8\%& 2.8\%& 2.8\%& 2.8\%& 2.8\%\\
% Sovereign bonds, &0.995& 2.8\% & 5.6\%& 5.6\%& 5.6\%& 2.8\%& 2.8\%\\
% long position&0.999 &8.3\% &11.1\%& 11.1\%& 11.1\%& 8.3\%& 11.1\%\\
% \\
% \hline \\
% Portfolio B:&0.99& 3.4\%& 3.4\%& 2.2\%& 2.2\%& 3.4\%& 3.4\%\\
% Corporate bonds,&0.995& 5.6\%& 5.6\%& 5.6\%& 5.6\%& 5.6\%& 5.6\%\\
% long position&0.999 &14.6\%& 14.6\%& 16.9\%& 16.9\%& 15.7\%& 16.9\%\\
% \\
% \hline \\
% Portfolio C:&0.99& 2.4\%& 2.4\%& 2.4\%& 2.4\%& 2.4\%& 2.4\%\\
% Combination of&0.995& 4.8\%& 4.8\%& 4.8\%& 4.8\%& 4.8\%& 4.8\%\\
% portfolios A and B&0.999& 12.8\%& 12.8\%& 13.6\%& 13.6\%& 12.8\%& 13.6\%\\ 
% \\
% \hline \\
% Portfolio D:&0.99& 0.0\%& 0.0\%& 0.0\%& 0.0\%& 0.0\% &0.0\%\\
% Corporate bonds,&0.995& 4.5\%& 4.5\%& 4.5\%& 0.0\%& 4.5\%& 4.5\%\\
% long/short position&0.999& 9.1\%& 13.6\% &9.1\% & 9.1\% & 13.6\% & 13.6\%\\
% \\ \hline
% \end{tabular}
% \end{table}

\begin{figure}
\centering
\includegraphics[width=0.8\textwidth]{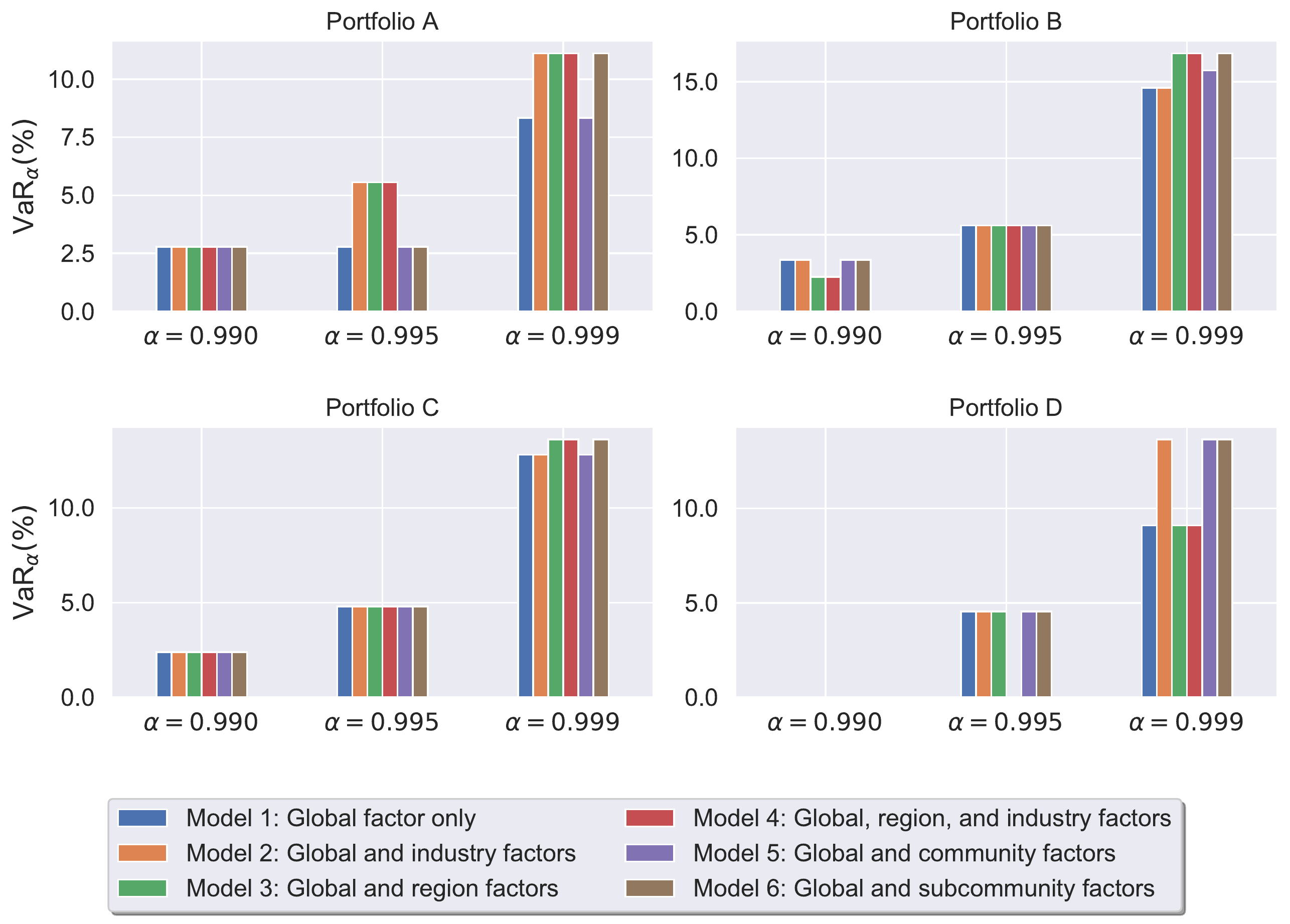}
\caption{Quantiles of the loss distributions obtained from the six model variants (global factor only, global/industry, global/region, global/industry/region, global/community, global/subcommunity factors) for the four synthetic test portfolios.}
\label{fig:risk_measures}
\end{figure}

Since both IRC in Basel 2.5 and DRC in FRTB are calculated based on a 99.9\% VaR over a capital horizon of one year, we rely on this risk measure in order to compare the impact of different correlation model configurations on portfolio risk. We calculate $\mbox{VaR}_{\alpha}(L)$ for selected confidence levels $\alpha\in \lbrace 0.99, 0.995, 0.999\rbrace$, including the 0.999 which corresponds to DRC, for each of the six model configurations and the synthetic test portfolios. The results are illustrated in \Cref{tab:losses} and \Cref{fig:risk_measures}.

For Portfolio A, consisting of long positions on sovereign issuers, Model 1 and Model 5 yield a DRC figure of 8.3\%, while the other model variants yield a slightly more conservative figure at 11.1\%. For Porfolio B, consisting of long positions on corporates, the values are higher and more variable, with Model 1 and Model 2 producing a DRC at 14.6\%, Model 5 at 15.7\%, and the rest of the models at 16.9\%. For the more diversified Portfolio C, models 1,2, and 5 produce a DRC of 12.8\%, while models 3,4,6 produce slightly higher figures at 13.6\%. Finally, for the long/short Portfolio D, models 1,3, and 4 yield a DRC of 9.1\%, while models 2,5, and 6 yield a more conservative 13.6\%.

In general, different model variants produce less variable losses for lower quantiles. Furthermore, although it is not straightforward to draw solid conclusions on one model being consistently more conservative, it seems that model based on global and subcommunity factors (Model 6) is among the most conservative models for all four portfolios, while the one-factor model (Model 1) is consistently among the least conservative. In addition, the DRC values produced by models 3 and 4 are in agreement for all the portfolios.

Another noteworthy observation has to do with the tails of the generated loss distributions. By analysing the relationship between the quantiles presented in \Cref{tab:losses}, it can be seen that all model variants seem to produce distributions with heavier tails than the standard normal distribution. For instance, the ratio between the 0.999 and the 0.99 quantile for Portfolio A is 2.96 for Model 1 and Model 5, and 3.96 for all the other model variants, compared to 1.33 for the standard normal distribution. Similar tail behaviour can be observed for all the portfolios.

\section{Concluding remarks}\label{sec:conclusions}

One of the most challenging problems in the study of complex systems is that of identifying the mesoscopic organisation of the constituting units. This amounts to detecting groups of units which are more densely connected internally than with the rest of the system. In case the units are represented by time series, a common approach is to regard correlation matrices as weighted networks and employ standard network community detection methods. Since such an approach can introduce biases, in this paper we adopt a principled approach based on Random Matrix Theory, leading to an algorithm that is able to identify internally correlated and mutually anti-correlated communities, in a multiresolution fashion.

Our methods are applied to the analysis of CDS time series with the aim of identifying mesoscopic groups of issuers whose similarities cannot be traced back to industry sectors or geographical regions. We use ten years of data including the stressed period between 2007 and 2009. The analysis reveals several interesting results with regards to the community structure. In addition, our results show that different time resolutions yield similar community structures and that these structures are stable over time; this renders the obtained communities useful for risk models.

Based on the detected communities, we derive factors and build a model for portfolio credit risk that is in line with regulatory requirements for the calculation of DRC. This model is then compared with industry-standard models based on global, country, and region factors. The models based on global, communities, and subcommunities factors provide a better fit to the data and lower error between the model-implied and the empirical pairwise correlations compared to the other two-factor models. 

To further explore the properties of the obtained factor models we set up four synthetic portfolios and generate loss distributions via Monte Carlo simulations. The results show that the model based on global and subcommunity factors is consistently among the most conservative. As a more general observation, all models produce distributions with heavy tails and the variability is higher in the higher quantiles, including the 0.999 which is of particular interest for DRC.

The present work emphasises that complexity-inspired models, in combination with more traditional modelling and simulation frameworks, can open up new avenues for risk management and fill some of the gaps and inadequacies of regulatory frameworks revealed by the global financial crisis. Our paper represents a first step towards incorporating community detection in risk models that can be used in a realistic set up. Further research could usefully explore applications in portfolio optimisation and hedging of credit sensitive instruments. An interesting avenue for further study that would establish the value of our findings for portfolio construction could be a rigorous analysis of the diversification potential for portfolios of risky instruments across communities and subcommunities, similar to the one conducted by \cite{aretz2013common} for country and industry factors.

\section*{Disclosure}
The opinions expressed in this work are solely those of the authors and do not represent in any way those of their current and past employers. No potential conflict of interest was reported by the authors.
\section*{Funding}
This project has received funding from the European Union’s Horizon 2020 research and innovation programme under the Marie Skłodowska-Curie Grant Agreement no. 675044 (\href{http://bigdatafinance.eu/}{http://bigdatafinance.eu/}), Training for Big Data in Financial Research and Risk Management. D. G. acknowledges support from the Dutch Econophysics Foundation (Stichting Econophysics, Leiden, the Netherlands)). D. G. and T. S. acknowledge support from the European Union’s Horizon 2020 research and innovation programme under grant agreement no. 871042 (SoBigData++), European Integrated Infrastructure for Social Mining and Big Data Analytics.

\bibliographystyle{rQUF}  
\bibliography{rQUFguide}

\appendix

\newpage

\section{Full regression results on the basis of \Cref{eq:regression1}}\label{app:reg_results}

This appendix presents the full regression results for the time-series regression models that are presented in \Cref{eq:regression1} of the main article. 

\begin{table}[h]
\begin{center}
\begin{minipage}{\textwidth}
\tbl{This table provides the results of the industry, region, community, and subcommunity factor analysis derived from CDS spread returns. The analysis is based on non-overlapping monthly log-returns and conducted over the period January 2007 through December 2016. After standardisation the CDS returns are used to derive global, industry, region and community factors as cross-sectional averages at each time point. The industry, region, and community factors are regressed onto the global factor; coefficients ($\gamma_I$, $\gamma_R$, $\gamma_C$, and $\gamma_{SC}$ respectively) as well as $R^2$ are provided in the table, complemented by the
standard deviation of the residual returns. $t$-tests are conducted to evaluate whether the coefficients differ from one.}
{\begin{tabular}{@{}crrrrrr}\toprule
Industry                    & $\gamma_I$ & $t$-statistic & $p$-value        & $R^{2}$ & $\sigma_I$ & $\sigma_G$ \\
\colrule
Basic Materials	&	1.027	&	0.993	&	32.3\%	&	92.3\%	&	20.1\%	&	\multirow{11}{*}{67.9\%} 	\\
Consumer Goods	&	0.988	&	-0.586	&	55.9\%	&	95.2\%	&	15.1\%	&		\\
Consumer Services	&	0.962	&	-1.676	&	9.6\%	&	93.7\%	&	16.9\%	&		\\
Energy   	&	1.021	&	0.491	&	62.5\%	&	82.7\%	&	31.7\%	&		\\
Financials	&	1.019	&	0.603	&	54.8\%	&	90.1\%	&	23.0\%	&		\\
Government	&	1.042	&	1.035	&	30.3\%	&	84.6\%	&	30.2\%	&		\\
Health Care	&	0.880	&	-3.306	&	0.1\%	&	83.2\%	&	26.9\%	&		\\
Industrials	&	1.028	&	1.493	&	13.8\%	&	96.3\%	&	13.6\%	&		\\
Technology	&	0.890	&	-4.178	&	0.0\%	&	90.7\%	&	19.4\%	&		\\
Telecommunications Services	&	1.023	&	0.982	&	32.8\%	&	94.0\%	&	17.5\%	&		\\
Utilities	&	1.009	&	0.350	&	72.7\%	&	93.2\%	&	18.5\%	&		\\
\colrule
Region                      & $\gamma_R$ & $t$-statistic & $p$-value        & $R^2$ & $\sigma_R$ & $\sigma_G$ \\
\colrule
Africa      	&	1.056	&	0.936	&	35.1\%	&	72.8\%	&	43.8\%	&	\multirow{9}{*}{67.9\%} 	\\
Asia        	&	1.115	&	2.883	&	0.5\%	&	86.9\%	&	29.3\%	&		\\
Eastern Europe	&	1.065	&	1.056	&	29.3\%	&	71.8\%	&	45.3\%	&		\\
Europe      	&	1.021	&	0.928	&	35.5\%	&	94.5\%	&	16.7\%	&		\\
India       	&	1.131	&	1.792	&	7.6\%	&	67.1\%	&	53.8\%	&		\\
Latin America	&	1.072	&	1.221	&	22.4\%	&	73.6\%	&	43.6\%	&		\\
Middle East	&	0.928	&	-1.282	&	20.2\%	&	69.7\%	&	41.6\%	&		\\
North America 	&	0.926	&	-3.702	&	0.0\%	&	94.9\%	&	14.6\%	&		\\
Oceania     	&	1.104	&	2.688	&	0.8\%	&	87.3\%	&	28.7\%	&		\\
\colrule
Community                   & $\gamma_C$ & $t$-statistic & $p$-value        & $R^{2}$ & $\sigma_C$ & $\sigma_G$\\
\colrule
A	&	0.967	&	-1.196	&	23.4\%	&	91.5\%	&	20.1\%	&	\multirow{4}{*}{67.9\%} 	\\
B	&	0.987	&	-0.526	&	60.0\%	&	92.6\%	&	18.9\%	&		\\
C	&	1.061	&	1.852	&	6.6\%	&	89.9\%	&	24.2\%	&		\\
D	&	0.989	&	-0.220	&	82.6\%	&	76.8\%	&	36.9\%	&		\\
\colrule
Subcommunity                   & $\gamma_{S}$ & $t$-statistic & $p$-value        & $R^{2}$ & $\sigma_S$ & $\sigma_G$\\
\colrule
A1	&	0.943	&	-1.665	&	0.098	&	86.5\%	&	25.3\%	&	\multirow{15}{*}{67.9\%} 	\\
A2	&	0.934	&	-2.052	&	0.042	&	87.6\%	&	23.9\%	&		\\
A3	&	1.093	&	2.260	&	0.026	&	85.6\%	&	30.4\%	&		\\
A4	&	0.999	&	-0.015	&	0.988	&	76.5\%	&	37.5\%	&		\\
B1	&	0.952	&	-1.516	&	0.132	&	88.5\%	&	23.2\%	&		\\
B2	&	1.067	&	2.267	&	0.025	&	91.8\%	&	21.7\%	&		\\
C1	&	1.022	&	0.474	&	0.636	&	80.4\%	&	34.3\%	&		\\
C2	&	0.995	&	-0.141	&	0.888	&	88.4\%	&	24.5\%	&		\\
C3	&	1.154	&	3.665	&	0.000	&	86.5\%	&	31.0\%	&		\\
D1	&	1.026	&	0.379	&	0.705	&	65.5\%	&	50.5\%	&		\\
D2	&	0.895	&	-3.202	&	0.002	&	86.4\%	&	24.1\%	&		\\
D3	&	0.980	&	-0.295	&	0.768	&	62.8\%	&	51.2\%	&		\\
D4	&	1.123	&	1.664	&	0.099	&	66.2\%	&	54.6\%	&		\\
D5	&	1.081	&	1.122	&	0.264	&	65.7\%	&	53.0\%	&		\\
D6	&	1.011	&	0.168	&	0.867	&	65.0\%	&	50.4\%	&		\\
\botrule
\end{tabular}}
\label{tab:results}
\end{minipage}
\end{center}
\end{table}

\newpage

\section{Empirical pairwise correlations}\label{sec:emp_correlations}

\begin{figure}[h!]
\centering
\includegraphics[width=0.85\textwidth]{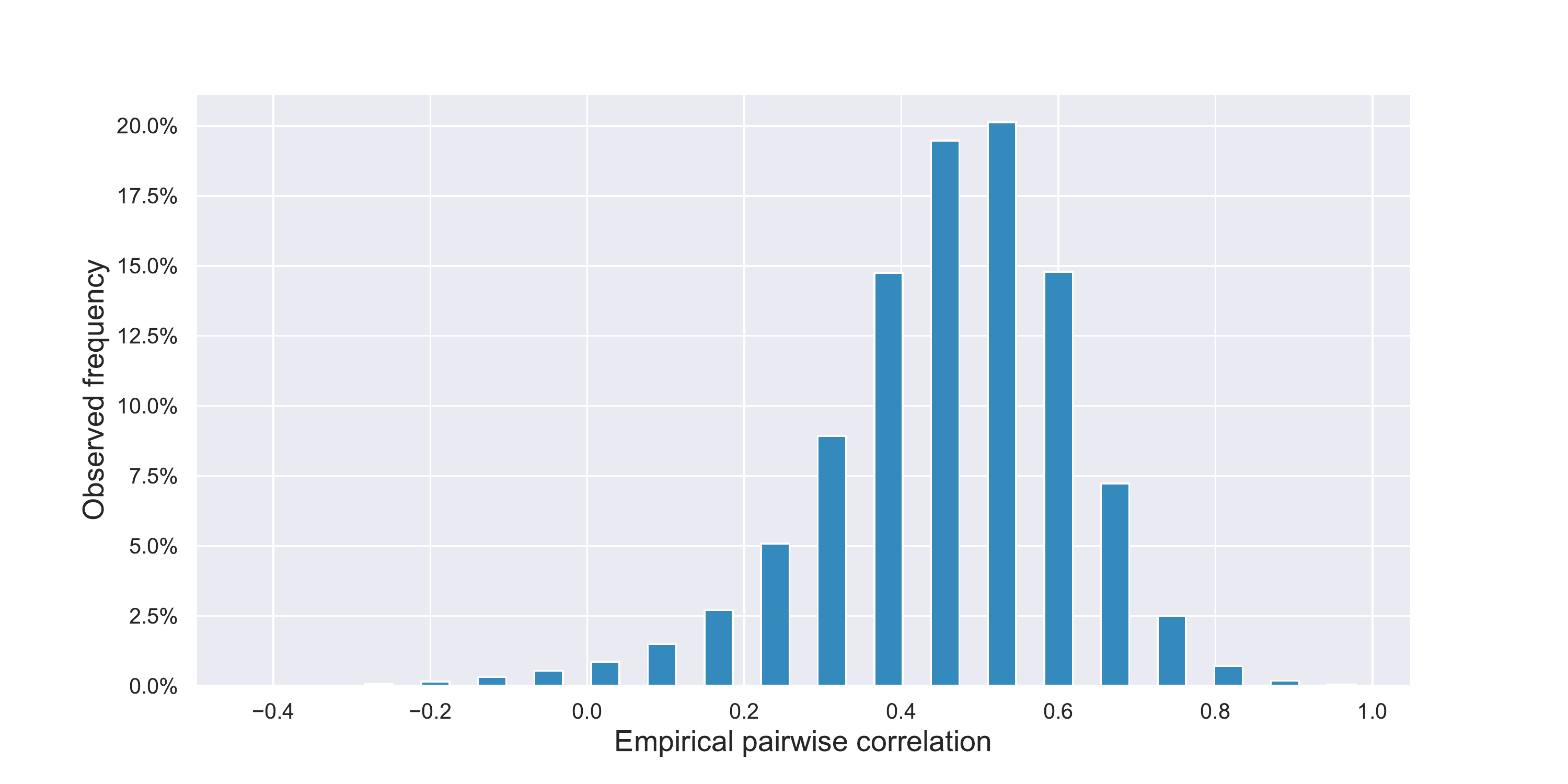}
\caption{Empirical pairwise correlation. The figure shows the distribution of the CDS spread correlation during the period January 2007 through
December 2016, based on Markit data for 786 issuers. The return correlation is calculated from non-overlapping monthly log-returns.}
\label{fig:emp_cor}
\end{figure}

\end{document}